\documentclass[longauth]{aa} 
\usepackage{graphicx}
\usepackage{txfonts}
\def\lofar{{LOFAR}}
\def\spitzer{{\em Spitzer}}

\def\chandra{{\em Chandra}}

\usepackage{amssymb}
\usepackage{mathtools}
\usepackage{natbib}
\usepackage{graphicx,url,multicol,enumitem,sidecap}
\usepackage[margin=20pt]{subfig} 
\usepackage{multirow} 

\def\msun{$M_{\odot}$}

\usepackage{color}

\begin{document} 

\title{Low-frequency radio absorption in Cassiopeia A}
\author{
M.~Arias\inst{1}\and 
J.~Vink\inst{1,38,39}\and 
F.~de Gasperin\inst{2,6}\and 
P.~Salas\inst{2}\and 
J.~B.~R.~Oonk \inst{2,6}\and 
R.~J.~van Weeren \inst{2}\and 
A.~S.~van Amesfoort \inst{6}\and
J.~Anderson\inst{3}\and 
R.~Beck\inst{4}\and 
M.~E.~Bell\inst{5}\and 
M.~J.~Bentum\inst{6,7}\and 
P.~Best\inst{8}\and 
R.~Blaauw\inst{6}\and 
F.~Breitling\inst{9}\and 
J.~W.~Broderick\inst{6}\and 
W.~N.~Brouw\inst{6,10}\and 
M.~Br\"uggen\inst{11}\and 
H.~R.~Butcher\inst{12}\and 
B.~Ciardi\inst{13}\and 
E.~de Geus\inst{6,14}\and 
A.~Deller\inst{15,6}\and 
P. ~C.~G.~van Dijk\inst{6}\and 
S.~Duscha\inst{6}\and 
J.~Eisl\"offel\inst{16}\and 
M.~A.~Garrett\inst{17,2}\and 
J.~M.~Grie\ss{}meier\inst{18,19}\and 
A.~W.~Gunst\inst{6}\and 
M.~P.~van Haarlem\inst{6}\and 
G.~Heald\inst{20,10,6}\and 
J.~Hessels\inst{1,6}\and
J.~H\"orandel\inst{21}\and 
H.~A.~Holties\inst{6}\and 
A.J.~van der Horst\inst{22}\and 
M.~Iacobelli\inst{6}\and 
E.~Juette\inst{23}\and 
A.~Krankowski\inst{24}\and 
J.~van Leeuwen\inst{6,1}\and 
G.~Mann\inst{9}\and 
D.~McKay-Bukowski\inst{25,26}\and 
J.~P.~McKean\inst{6,10}\and 
H.~Mulder\inst{6}\and 
A.~Nelles\inst{27}\and 
E.~Orru\inst{6}\and 
H.~Paas\inst{28}\and 
M.~Pandey-Pommier\inst{29}\and 
V.~N.~Pandey\inst{6,10}\and 
R.~Pekal\inst{30}\and 
R.~Pizzo\inst{6}\and 
A.~G.~Polatidis\inst{6}\and 
W.~Reich\inst{4}\and 
H.~J.~A.~R\"ottgering\inst{2}\and 
H. Rothkaehl\inst{31}\and 
D.~J.~Schwarz\inst{32}\and 
O.~Smirnov\inst{33,34}\and 
M.~Soida\inst{35}\and 
M.~Steinmetz\inst{9}\and 
M.~Tagger\inst{18}\and 
S.~Thoudam\inst{36}\and 
M.~C.~Toribio\inst{2,6}\and 
C. Vocks\inst{9}\and 
M.~H.~D.~van der Wiel\inst{6}\and 
R.~A.~M.~J.~Wijers\inst{1} \and
O.~Wucknitz\inst{4}\and 
P.~Zarka\inst{37}\and 
P.~Zucca\inst{6}
}

\institute{
Anton Pannekoek Institute for Astronomy, University of Amsterdam, Science Park 904, 1098 XH Amsterdam, The Netherlands \and
Leiden Observatory, Leiden University, PO Box 9513, 2300 RA Leiden, The Netherlands \and
Helmholtz-Zentrum Potsdam, GFZ, Department 1: Geodesy and Remote Sensing, Telegrafenberg, A17, 14473 Potsdam, Germany \and
Max Planck Institute for Radio Astronomy, Auf dem H\"ugel 69, 53121 Bonn, Germany \and
University of Technology Sydney, 15 Broadway, Ultimo NSW 2007, Australia \and
ASTRON, Netherlands Institute for Radio Astronomy, Postbus 2, 7990 AA, Dwingeloo, The Netherlands \and
Eindhoven University of Technology, P.O. Box 513, 5600 MB  Eindhoven, The Netherlands \and
Institute for Astronomy, University of Edinburgh, Royal Observatory of Edinburgh, Blackford Hill, Edinburgh EH9 3HJ, UK \and
Leibniz-Institut f\"{u}r Astrophysik Potsdam (AIP), An der Sternwarte 16, 14482 Potsdam, Germany \and
Kapteyn Astronomical Institute, PO Box 800, 9700 AV Groningen, The Netherlands \and
University of Hamburg, Gojenbergsweg 112, 21029 Hamburg, Germany \and
Research School of Astronomy and Astrophysics, Australian National University, Canberra, ACT 2611 Australia \and
Max Planck Institute for Astrophysics, Karl Schwarzschild Str. 1, 85741 Garching, Germany \and
SmarterVision BV, Oostersingel 5, 9401 JX Assen, The Netherlands \and
Centre for Astrophysics \& Supercomputing, Swinburne University of Technology John St, Hawthorn VIC 3122 Australia \and
Th\"{u}ringer Landessternwarte, Sternwarte 5, 07778 Tautenburg, Germany \and
Jodrell Bank Center for Astrophysics, School of Physics and Astronomy, The University of Manchester, Manchester M13 9PL,UK \and
LPC2E - Universite d'Orleans/CNRS, France \and
Station de Radioastronomie de Nancay, Observatoire de Paris - CNRS/INSU, USR 704 - Univ. Orleans, OSUC , route de Souesmes, 18330 Nancay, France \and
CSIRO Astronomy and Space Science, 26 Dick Perry Avenue, Kensington, WA 6151, Australia  \and
Department of Astrophysics, Radboud University Nijmegen, P.O. Box 9010, 6500 GL Nijmegen, The Netherlands \and
Department of Physics, The George Washington University, 725 21st Street NW, Washington, DC 20052, USA \and
Astronomisches Institut der Ruhr-Universit\"{a}t Bochum, Universitaetsstrasse 150, 44780 Bochum, Germany \and
University of Warmia and Mazury in Olsztyn, Oczapowskiego 1, 10-957 Olsztyn, Poland \and
Department of Physics and Technology, University of Troms\o, Norway \and
STFC Rutherford Appleton Laboratory,  Harwell Science and Innovation Campus,  Didcot  OX11 0QX, UK \and
Department of Physics and Astronomy, University of California Irvine, Irvine, CA 92697, USA \and
Center for Information Technology (CIT), University of Groningen, The Netherlands \and
Centre de Recherche Astrophysique de Lyon, Observatoire de Lyon, 9 av Charles Andr\'{e}, 69561 Saint Genis Laval Cedex, France \and
Poznan Supercomputing and Networking Center (PCSS) Poznan, Poland \and
Space Research Center PAS, Bartycka 18 A, 00-716, Warsaw, Poland, \and
Fakult\"{a}t f\"{u}r Physik, Universit\"{a}t Bielefeld, Postfach 100131, 33501, Bielefeld, Germany \and
Department of Physics and Electronics, Rhodes University, PO Box 94, Grahamstown 6140, South Africa \and
SKA South Africa, 3rd Floor, The Park, Park Road, Pinelands, 7405, South Africa \and
Jagiellonian University, Astronomical Observatory, Orla 171, 30-244 Krakow, Poland \and
Department of Physics and Electrical Engineering, Linnaeus University 35195, Vaexjoe, Sweden \and
LESIA \& USN, Observatoire de Paris, CNRS, Place J. Janssen, 92195 Meudon, France \and 
GRAPPA, University of Amsterdam, Science Park 904, 1098 XH Amsterdam \and SRON, Netherlands Institute for Space Research, Utrecht
}

\date{Received 2/12/2017; accepted 13/1/2018}

 
  \abstract
   {
   Cassiopeia A is one of the best-studied supernova remnants. Its bright radio and X-ray emission
   is due to shocked ejecta. Cas A is rather unique in that the unshocked ejecta can also be studied: through emission in the infrared,
   the radio-active decay of $^{44}$Ti, and the low-frequency free-free absorption caused by cold ionised
   gas, which is the topic of this paper.}
   {Free-free absorption processes are affected by the mass, geometry, temperature, and ionisation conditions in the absorbing gas. Observations at the lowest radio frequencies can constrain a combination of these  properties. }
   {We used Low Frequency Array (\lofar)\ Low Band Antenna observations at 30--77 MHz and Very Large Array (VLA) L-band observations at 1--2 GHz to fit for internal absorption as parametrised by the emission measure. 
   We simultaneously fit multiple UV-matched images with a common resolution of 17\arcsec (this corresponds to 0.25 pc for a source at the distance of Cas A).
   The ample frequency coverage allows us 
separate the relative contributions from the absorbing gas, the unabsorbed front of the shell, and the absorbed back of the shell to the emission spectrum.
   We explored the effects that a temperature lower than the $\sim$100--500 K proposed from infrared observations and a high degree of clumping can have on 
   the derived physical properties of the unshocked material, such as its mass and density. We also compiled integrated radio flux density measurements, fit for the absorption processes that occur in the radio band, and considered their effect on the secular decline of the source.}
   {We find a mass in the unshocked ejecta of $M = 2.95 \pm {0.48}$ \msun\ for an assumed gas temperature of $T=100$~K. This estimate is reduced for colder  gas temperatures and, most significantly, if the ejecta are clumped. We measure the reverse shock to have a radius of $114$\arcsec $\pm $6\arcsec  and be centred at 23:23:26, +58:48:54 (J2000). We also find that a decrease in the amount of mass in the unshocked ejecta (as more and more material meets the reverse shock and heats up) cannot account for the observed low-frequency behaviour of the secular decline rate.}
   {To reconcile our low-frequency absorption measurements with models that reproduce much of the observed behaviour in Cas A and predict little mass in the unshocked ejecta, the ejecta need to be very clumped or the temperature in the cold gas needs to be low ($\sim10$~K). Both of these options are plausible
   and can together contribute to the high absorption value that we find.}


   \maketitle
%

\section{Introduction}

Supernova remnants (SNRs) are characterised by radio synchrotron  spectra with relatively steep indices \citep[$\alpha \approx 0.5$,][]{dubner15}, compared
to pulsar wind nebulae and HII regions ($\alpha\approx 0.25$ and $\alpha\approx 0.1$ respectively; $S \propto \nu^{-\alpha}$).  
As a result, SNRs are bright at low frequencies, which makes them excellent
targets for low-frequency radio telescopes. In this regime, however, the approximation of a power-law-shaped spectrum may not hold, as free-free absorption
effects from the  cold but partially ionised interstellar medium become important. 

The effect of interstellar absorption has a clear imprint on the radio spectrum of the brightest SNR, Cassiopeia A \citep[Cas A,][]{baars77}, the remnant of a supernova
that must have occurred around 1672 \citep{thorstensen01}. \citet{kassim95} discovered that 
the spectrum of Cas A is affected by absorption from cold, unshocked ejecta internal to the shell of Cas A,
in addition to interstellar free-free absorption. 
These ejecta cooled as a result of adiabatic expansion and have yet to encounter the reverse shock and reheat.
The unshocked ejecta are usually difficult to study, since the radiative output of SNRs is dominated by the contribution from the shocked ambient medium and ejecta
emitting synchrotron in the radio (and sometimes up to X-rays), collisionally heated  dust emission, and thermal  X-ray emission.

Internal absorption therefore provides a means for studying an important but elusive component of the SNR. Cas A is rather unique in that the
unshocked ejecta are also associated with infrared line and dust emission \citep{ennis06,smith09,isensee10,eriksen09,delaney10,delooze17}. 
Moreover, some of the radioactive
line emission in X-rays and gamma-rays can come from unshocked material, since radioactive decay does not depend on whether the material is shocked. This is the case for $^{44}$Ti \citep{iyudin94,vink01a,renaud06,grefenstette16}.

In addition to free-free absorption, the low-frequency spectrum of synchrotron sources can be affected by synchrotron self-absorption. This is usually only important for compact radio sources with high magnetic fields, but \citet{atoyan00} suggested that the high magnetic
 fields in the shell of Cas A may also give rise to synchrotron self absorption in compact, bright knots. The magnetic fields in the shell have been
 estimated to be as high as $B\sim 1~$mG based on the minimum energy argument \citep[e.g.][]{rosenberg70}, whereas the combination of gamma-ray
 observations (which set an upper limit on the bremsstrahlung from relativistic electrons) and radio emission provide a lower limit to the average magnetic field of
 $B>0.12$~mG \citep{abdo10}.
 
 In order to study the various phenomena that may shape the morphology and spectrum of Cas A at low frequencies, we analysed data obtained with
 the Low Frequency Array \citep[\lofar,][]{vanhaarlem13}. \lofar\ consists of two separate arrays, a low-band antenna array (LBA) that covers the $10-90$~MHz range,
 and a high-band antenna array (HBA) that covers the $115-250$~MHz range. This study is based on LBA observations only. \lofar\ is a phased-array telescope:
 the beam allows for simultaneous pointings as it is digitally formed. It combines the ability to observe at the lowest frequencies accessible from Earth with ample bandwidth and with an angular resolution of 17 arcseconds at 30 MHz. In order to study the effects of absorption at low frequencies, we combine
 the \lofar\ data with recent L-band Very Large Array (VLA) data between 1 and 2 GHz.
 
\cite{delaney14}  estimated the mass and density in the unshocked ejecta from optical depth measurements. Our work extends their study to lower frequencies and a broader bandwidth. 
Based on our analysis, we provide a new mass estimate and discuss the systematic uncertainties associated with this value, most notably the important effects of
ejecta temperature and clumping.
We show
that the reverse shock is not as shifted with respect to the explosion centre as is indicated by X-ray studies  \citep{gotthelf01a,helder08}.
Finally, we discuss the contribution of internal free-free absorption to the integrated flux of Cas A, in particular to its secular evolution, 
as more and more unshocked ejecta are heated by the reverse shock.


\section{Data reduction}

\subsection{LOFAR observations}

The data were taken in August 2015 as a legacy data set part of the \lofar \, commissioning cycle. The beam was split, and the
low-frequency calibrator 3C380 was observed simultaneously with the source. In the case of a bright, well-studied source like Cas A, the calibrator is just used for (a) determining the quality of the ionosphere at any given time, and (b) rescaling the amplitudes. 

Since the data are intended to be a legacy data set of the brightest radio sources in the sky (Cas A, Cyg A, Tau A, and Vir A, collectively referred to as \lq A-team'), the full array was used, including core, remote, and international stations. We still lack a well-understood method
for combining the sensitivity to large-scale diffuse emission provided by the short baselines with the VLBI resolution of the international stations. 
For this reason, we ignored the international baselines and only analysed the Dutch configuration of the array, with baselines of up to 120 km.

Given the wide field of view of \lofar, particularly in the LBA, A-team sources can easily enter a side lobe and outshine entire fields. It is a standard practice to \textit{\textup{demix}} these sources, that is, to subtract their contribution to the visibilities in any given observation \citep{vandertol17}. Cygnus A was demixed from the Cas A data, and both Cas A and Cyg A were demixed from the calibrator. The data were further flagged and averaged down from high spectral resolution (64 channels per subband) to four channels per subband. The demixing, RFI flagging and averaging were done using the LOFAR GRID preprocessing pipeline (Oonk et al. in prep.) on the SURFsara GINA cluster that is part of the Dutch GRID infrastructure. The LOFAR software and pipelines for this infrastructure are developed and maintained by the LOFAR e-infra group \cite[Oonk et al. in prep.;][]{mechev17}.

The calibration entailed  removing the effects of the beam, the ionosphere, the clock differences, and the bandpass. The ionospheric delay and the differential Faraday rotation are strongly frequency dependent ($\propto 1/\nu $ and $\propto 1/\nu^2$, respectively), and so the calibration was carried out channel by channel. 
The source was calibrated against a 69 MHz model from 2011 observations of Cas A, as referenced in Fig. 1 of \cite{oonk17}.

\begin{figure*}
   \centering
         \includegraphics[width=\textwidth]{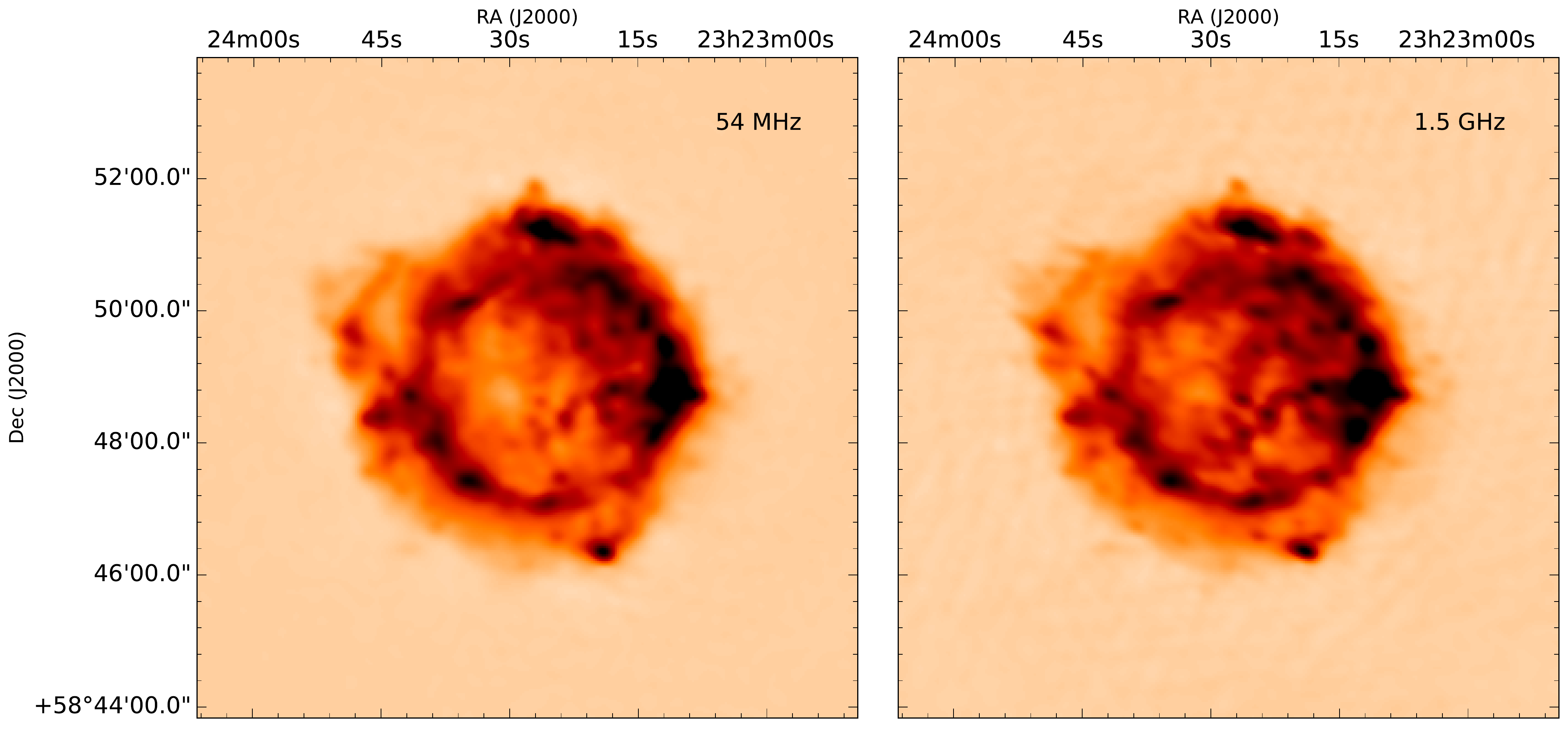}
      \caption{
     \textit{Left:} Cas A in the \textit{LOFAR} LBA. The central frequency is 54 MHz, the beam size is 10\arcsec, the noise is 10 mJy/beam, and the dynamic range is $13\,000$.
     \textit{Right:}
     Cas A in the VLA L band. Continuum image from combining the spectral windows at 1378 and 1750 MHz. The resolution is 14\arcsec 
     $\times$ 8\arcsec with a position angle of $70^ \mathrm{o}$, and the noise is 17 mJy/beam.  
              }
         \label{cas_a_lba}
   \end{figure*}

The visibilities were imaged with the \texttt{wsclean} software \cite[]{offringa14}, in the multiscale setting and with a Briggs parameter of $-0.5$. Several iterations of self-calibration against the clean components model were performed to make Fig.~\ref{cas_a_lba} (left).

\subsection{Narrow-band images}

Although the total bandwidth of the \lofar \, LBA configuration employed during these observations is continuous from 30 to 77 MHz, the signal-to-noise ratio in the case of Cas A is so high that it is possible to make narrow bandwidth (i.e., $\sim 1$ MHz) images in order to study the spectral behaviour of specific regions within the remnant. For the analysis presented here we made images at 30, 35, 40, 45, 50, 55, 60, 65, 71, and 77 MHz. In order to sample the same spatial scales and have images of a common angular resolution, 
all  images were made with a $u-v$ range of 500 to $12\,000 \, \lambda$. This corresponds to scales of 7 arcmin to 17 arcsec for a source the size of Cas A 
($\sim 5$ arcmin). The images are presented in Appendix A.

The in-band spectral behaviour of the LBA is not yet well understood. 
Hence, we bootstrapped the total flux densities of 
 the narrow-band images  (in a masked region containing Cas A)  to $S_\nu = S_{1\rm{GHz}} \left( \frac{\nu}{1\, \rm{GHz}} \right)^{-\alpha}$, with $ S_{1\rm{GHz}}=2720$ Jy and $\alpha=0.77$ \citep{baars77}. 

The flux density per pixel was measured and fitted for internal free-free absorption in the manner described in Sect.~\ref{em_det}. For the purposes of this study, we are concerned with the relative variations of flux in different locations of the remnant. Moreover, the lowest frequency image of 30 MHz is above the turnover in integrated spectrum of Cas A at 20 MHz, and the data were all taken simultaneously. This means that the usual issues that make it difficult to compare Cas A images (expansion, a time- and frequency-varying secular decline, and lack of data points at low radio frequencies) do not affect the results of our analysis.

\subsection{Spectral index map}

\begin{figure*}
   \centering
   \includegraphics[width=\columnwidth]{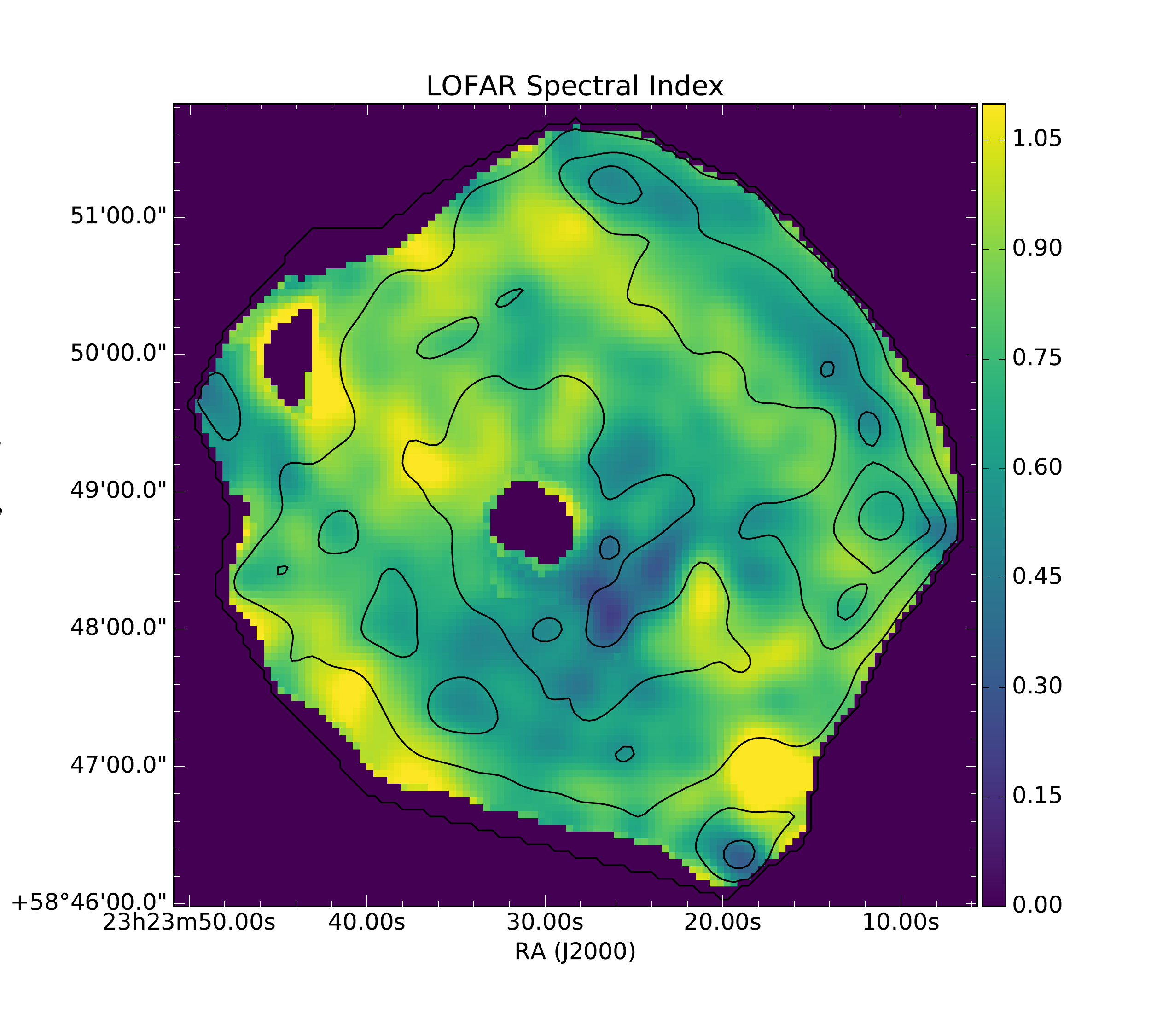}
      \includegraphics[width=\columnwidth]{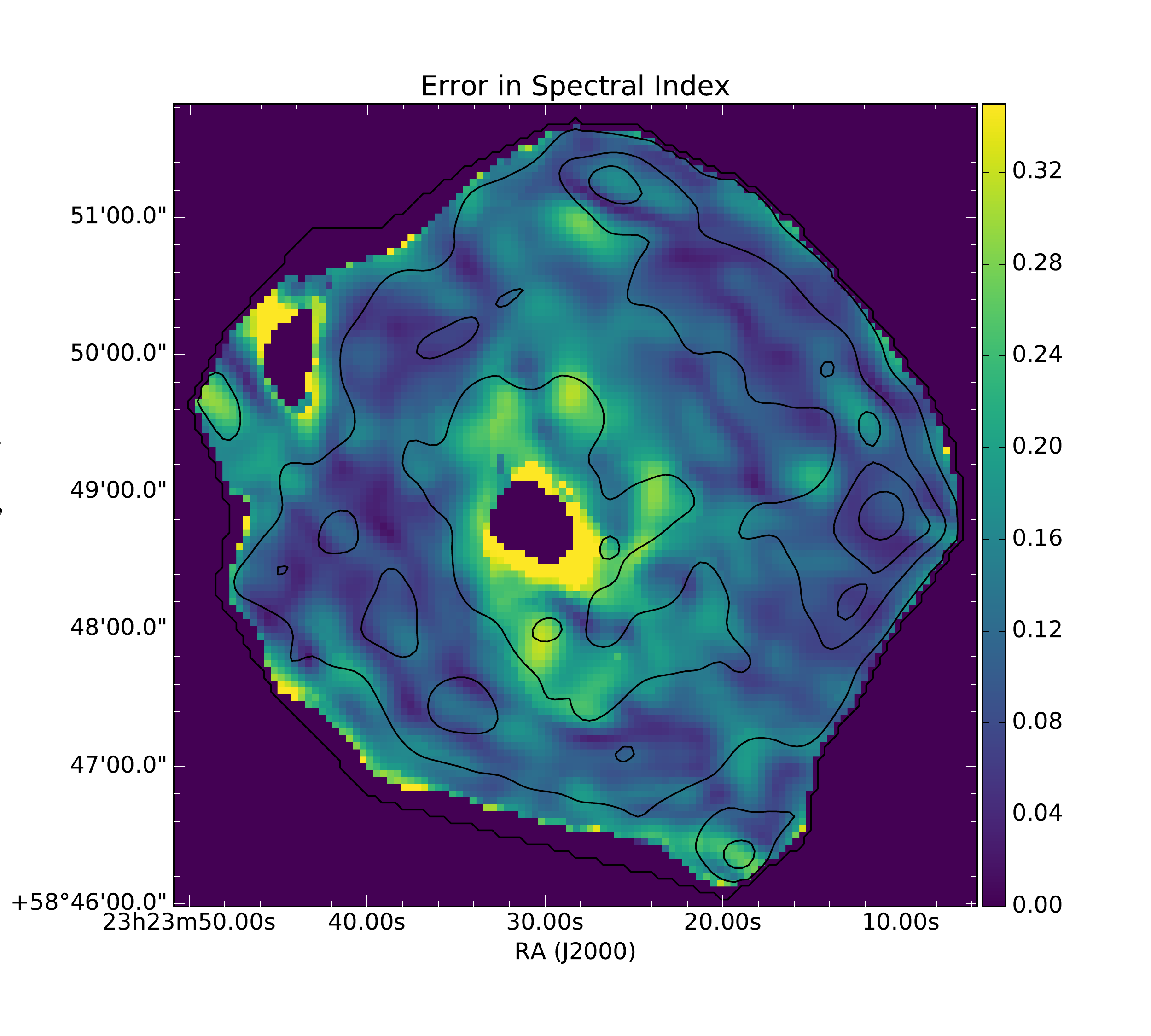}
      \caption{
     \textit{Left:} spectral index map made from fitting a power law to all the narrow-band \lofar \, images. Each image had a 10~$\sigma$ lower cut.
     \textit{Right:} square root of the diagonal element of the covariance matrix of the fit corresponding to $\alpha$.
     Overlaid are the radio contours at 70 MHz.              }
         \label{spx}
   \end{figure*}

We made a spectral index map from all the narrow-band \lofar\ images. Pixels with less than ten times the background rms flux density were set to zero for each map. 
We fitted a power law (i.e., amplitude and spectral index) for each pixel for which at least four images made the 10 $\sigma$ cut.
The best-fit values of $\alpha$ are shown in Fig. \ref{spx} (left). Figure \ref{spx} (right) shows the square root of the diagonal element of the covariance matrix corresponding to $\alpha$. We
note that because we bootstrapped the total flux density
for each individual map to the \citet{baars77} flux scale, the brightness-weighted average spectral index in Fig. \ref{spx} is by definition $\alpha=0.77$.

\cite{delaney14} presented a spectral index map between 74 MHz and 330 MHz. 
Our spectral map has features similar to theirs, particularly the centre-southwest region with a flatter index that they identify as the region of low-frequency absorption. 
However, our spread around the average value of $\alpha = 0.77$ is much larger than shown in the higher frequency map.  
It is expected that the map at lower frequencies would have more variance in the spectral index, since we probe lower frequencies that are
more sensitive to absorption. 

It is possible that some of the steeper gradients in the map are artefacts introduced by self-calibrating the individual images independently,
since iterations of self-calibration can result in small coordinate shifts. We tested whether our maps were affected by this effect
by  also making a spectral index map using a resolution a factor two lower, making it less sensitive to small coordinate shifts. However,
this did not alter the measured variation in spectral index

\subsection{VLA observations}

We observed Cassiopeia~A with the Karl G. Jansky Very Large Array \citep[VLA,][]{Thompson1980,Napier1983,Perley2011} during June and August 2017 (project $17\mbox{A--}283$).
The observations were carried out in the C array configuration using the L-band ($1\mbox{--}2$~GHz) receivers.
The correlator was set up to record data over $27$ spectral windows: three windows for the continuum, and the remaining over radio recombination and hydroxyl radical lines.
For each continuum window, we used a $128$~MHz bandwidth with $64$ spectral channels. 
These were centred at $1122$, $1378,$ and $1750$~MHz.
To determine the absolute flux density scale, we observed $3\mbox{C}286$ at the beginning of each observation.
We also used $3\mbox{C}286$ as bandpass calibrator.
As a phase reference, we used $3\mbox{C}468$.

The data reduction was performed using the Common Astronomy Software Applications package \citep[\emph{CASA},][]{McMullin2007}.
To calibrate the data, we used the VLA scripted calibration pipeline. 
The pipeline determines delay, bandpass, and complex gain solutions from the calibrator scans ($3\mbox{C}286$ and $3\mbox{C}468.1$) and applies them to the target data.
To image the continuum, we combined the calibrated data from different observations and used the multi-scale multi-frequency deconvolution implemented in \emph{CASA} \citep[e.g.][]{Rau2011}.
During the deconvolution, we used Gaussians with full-widths at half-maximum of $2\arcsec$, $6\arcsec$, $12\arcsec$, $24\arcsec,$ and $48\arcsec$, Briggs weighting with a robust parameter of $0$, and two terms for the Taylor series frequency dependence of the sky model.
The resulting continuum image (Fig.~\ref{cas_a_lba}, right) 
has a resolution of $14\arcsec\times8\arcsec$ with a position angle of $70^{\circ}$ and a noise of $17$~mJy/beam.

\subsection{Archival observations}

For our free-free absorption fit, we also made use of the VLA images of Cas A at 330 MHz and at 5 GHz in \cite{delaney14}. We smoothed these images to the resolution of the 30 MHz image and rescaled to the same pixel size of 3\arcsec. These images were also bootstrapped to the same frequency scale as the \lofar \, narrow-band images so as to ignore secular decline fading. The \cite{delaney14} images are  from a different epoch (data taken in 1997 and 2000, respectively). We neglected the effects of expansion, although it is measurable over the 18-year period. The expansion would correspond to $\sim 12''$ for the fastest moving optical knots, and to $\sim 2''$ for the radio-emitting material identified in \cite{anderson95}. This is smaller than the angular resolution of the maps used for our study ($3$~\arcsec\ pixel).

Finally, a note on the $u-v$ coverage of the VLA images. The 5 GHz image has a $u-v$ range of 500--81\,000 $\lambda$, and the 330 MHz has 700--81\,000 $\lambda$. The high $u-v$ cut affects the angular resolution of the image. The \lofar \, 500--12\,000 $\lambda$ images are not insensitive to flux on scales more compact than $12\,000 \lambda$, they just do not resolve it, and so
the flux densities per pixel can be compared safely if the high-resolution images are resolved down. The low $\lambda$ cuts of the \lofar \, and 5 GHz images are the same. In the case of the 330 MHz image, it does not probe scales of 500--700$ \lambda$ (i.e. 5\arcmin\
 to 7\arcmin), which the rest of the images do probe. This means that the narrow-band images may contain a low-level (very) diffuse \lq background' scale that is missed by the 330 MHz image. Therefore, we only take the flux density at 330 MHz to be a lower limit.


\section{Low-frequency map analysis}
\label{em_det}

\cite{delaney14} measured the low-frequency absorption using VLA data 
by comparing spectral index differences based on 330 MHz and 1.4 GHz maps, and 74 MHz and 330 MHz maps. 
They find that a region in the centre-west of the SNR displays spectral index flattening (a steeper value of the spectral index in the 330 MHz to 1.4 GHz map than in the 74 to 330 MHz map). 
They confirmed the suggestion by \citet{kassim95} that there is internal free-free absorption,
by finding a correlation between the regions of flat spectral index and the infrared emission from the unshocked ejecta as seen with \spitzer.
They concluded that both the IR emission and low-frequency absorption trace the same material.  

They argued that free-free absorption measurements of the optical depth coupled with assumptions about the geometry of the ejecta can yield an estimate of the mass in the unshocked ejecta. In this paper we follow a similar reasoning to that of \cite{delaney14}, but reach a different value of the mass in the unshocked ejecta from lower frequency data, a different analysis technique, and correcting for two misinterpreted parameters in their paper\footnote{They introduce the symbol $Z$ in their equation for the free-free optical depth as \lq the average atomic number of the ions dominating the cold ejecta' and take a value of 8.34, when in fact it is the average number of ionisations, and an 
$Z =$ 2--3 is more reasonable. 
They later estimate the density as the product of the number density of ions, the mass of the proton, and the average atomic number. This last value should be the average \textit{\textup{mass}} number.}. For the sake of clarity, we explicitly state all the relevant equations in this section.

One way in which our analysis method is different from the method
used in \cite{delaney14} is that we make use of \lofar 's multiwavelength capabilities by simultaneously fitting all images pixel by pixel, instead of comparing two spectral index maps.
This is more robust, as it uses the intrinsic spectral signatures of free-free absorption, and it is also less sensitive to small  artefacts in an individual image.

Our method is as follows:
\begin{enumerate}
\item Measure the flux density per $3\arcsec \times 3\arcsec$ pixel of the SNR images with spacings of 5 MHz.
\item For each pixel, fit for free-free absorption as parameterised by a factor of emission measure $EM$ (see below), number of ion charges $Z$, and temperature $T$.
\item Assuming a specific $T$ and $Z$, make an emission measure map.
\item Convert emission measure into a mass estimate by assuming a specific geometry.
\end{enumerate}

We illustrate the analysis that is performed per pixel by showing the fits to the region in the southwest of the remnant identified by \cite{delaney14} as the region of internal free-free absorption. The flux density in this region was measured in each image, and the plot in Fig.~\ref{abs_reg_fit} refers to these data points.

\subsection{Free-free absorption}

The coefficient for free-free absorption in the Rayleigh-Jeans approximation \cite[]{rohlfs09} is
\begin{equation}
\kappa_\nu = \frac{4 Z^2 e^6}{3c}\frac{n_\mathrm{e} n_\mathrm{i}}{\nu^2}\frac{1}{\sqrt{2 \pi (mkT)^3}} g_{\mathrm{ff}},
\end{equation}
where $c$ is the speed of light, $k$ is the Boltzmann constant, $Ze$ is the charge of the ion, $m$ is the mass of the electron, $n_\mathrm{e}$ and $n_\mathrm{i}$ are the number densities of electrons and ions, and $g_{\mathrm{ff}}$ is a Gaunt factor, given by
\begin{equation}
g_{\mathrm{ff}} = 
\begin{cases}
\ln\left[49.55 \, Z^{-1} \left(\frac{\nu}{\rm{MHz}}\right)^{-1} \right] + 1.5 \ln \frac{T}{\mathrm{K}}  \\  \\  1 & \hspace{-4cm} \text{for} \,\,\, \frac{\nu}{\rm{MHz}}>>\left(\frac{T}{\mathrm{K}}\right)^{3/2}.
\end{cases}
\end{equation}
The free-free optical depth is $\tau_\nu = - \int^{s_{\mathrm{out}}}_{s_{\mathrm{in}}} \kappa_\nu (s') ds'$. Integrating along the line of sight, using $\frac{n_\mathrm{i}}{n_\mathrm{e}} = \frac{1}{Z}$ and $EM  \equiv \int_{0}^{s} n_\mathrm{e}^2 ds'$ , and substituting for numerical values, we obtain the following equation for the free-free optical depth:
\begin{equation}
\tau_\nu = 3.014 \times 10^{4} \, Z \,\left( \frac{T}{\rm{K}} \right)^{-3/2} \left( \frac{\nu}{\rm{MHz}} \right)^{-2} \left( \frac{{EM}}{\rm{pc \,cm}^{-6}} \right) g_{\mathrm{ff}}.
\label{ff_tau}
\end{equation}

We recall that we used narrow-band images at a number of frequencies, and that their flux density has been bootstrapped to a power law with spectral index $\alpha = 0.77$. Since the flux densities were fixed to a power-law distribution, we do not need to account for the contribution of the interstellar medium (ISM) to low-frequency absorption in our fit procedure. We show later in the paper that this contribution is small even for our lowest frequency (30 MHz) image. 

The material responsible for internal absorption is \textit{\textup{inside}} the shell,  so that it can never absorb the front half of the shell. Hence, we can model the flux density  as\begin{equation}
S_\nu = (S_{\nu, \mathrm{front}} + S_{\nu, \mathrm{back}} \, e^{-\tau_{\nu, \mathrm{int}}}) \,e^{-\tau_{\nu, \mathrm{ISM}}}. 
\end{equation}

Cas A is quite clumpy, however, and this can affect the relative synchrotron brightness in the front and the back (consider if we could look at Cas A from the west; most of the emission would come from the bright knot in the west, and only the fainter eastern side of the shell would be absorbed by internal cold ejecta). We parameterise this by taking $S_{\nu, \mathrm{front}} = f S_\nu$, with $(1-f)$ the covering fraction of the absorbing material.

\subsection{Fitting}

The flux density in each pixel was measured per frequency (i.e. for each image), and fitted to the following equation:

\begin{equation}
S_\nu = S_0 \left( \frac{\nu}{\nu_0} \right)^{-\alpha} (f + (1-f)e ^{-\tau_{\nu, \mathrm{int}}}),
\label{fitting}
\end{equation}
where 
\begin{equation}
\tau_\nu = 3.014 \times 10^{4} \, \left( \frac{\nu}{\rm{MHz}} \right)^{-2} g_{\mathrm{ff}} (T=100 \, \mathrm{K}, Z=3) \, X(\nu, Z, T),
\label{fitted_equation}
\end{equation}
and 
 \begin{equation}
 X(\nu, Z, T) = Z \,\left( \frac{T}{\rm{K}} \right)^{-3/2} \left( \frac{{EM}}{\rm{pc \,cm}^{-6}} \right) \left( \frac{g_{\mathrm{ff}}(T, Z) }{g_{\mathrm{ff}} (T=100\, \mathrm{K}, Z = 3)} \right).
 \label{eq_x}
\end{equation}
We set $\alpha = 0.77$ \citep{baars77} and fitted for each pixel for $S_0$, $f$, and $X$ using the package non-linear least-squares minimisation and curve fitting for Python, \texttt{lmfit}. We
note that $X$ now contains all dependencies on the temperature and ionisation of the plasma. 
We took the rms pixel fluctuations using the background region for each map (that is, the regions not containing flux from Cas~A) as errors.

\begin{figure*}
\centering
\begin{multicols}{2}
\includegraphics[width=\columnwidth]{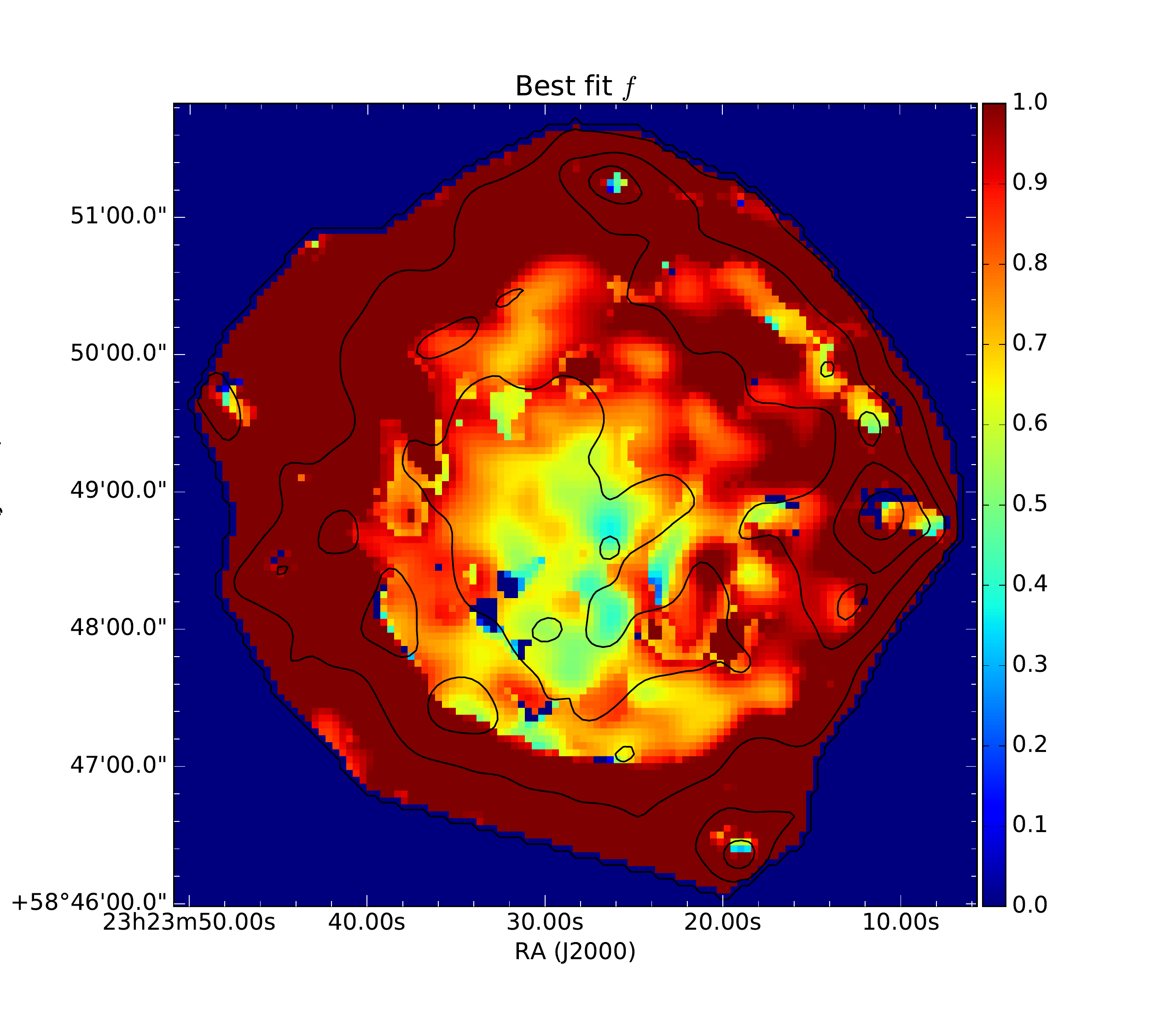}
\includegraphics[width=\columnwidth]{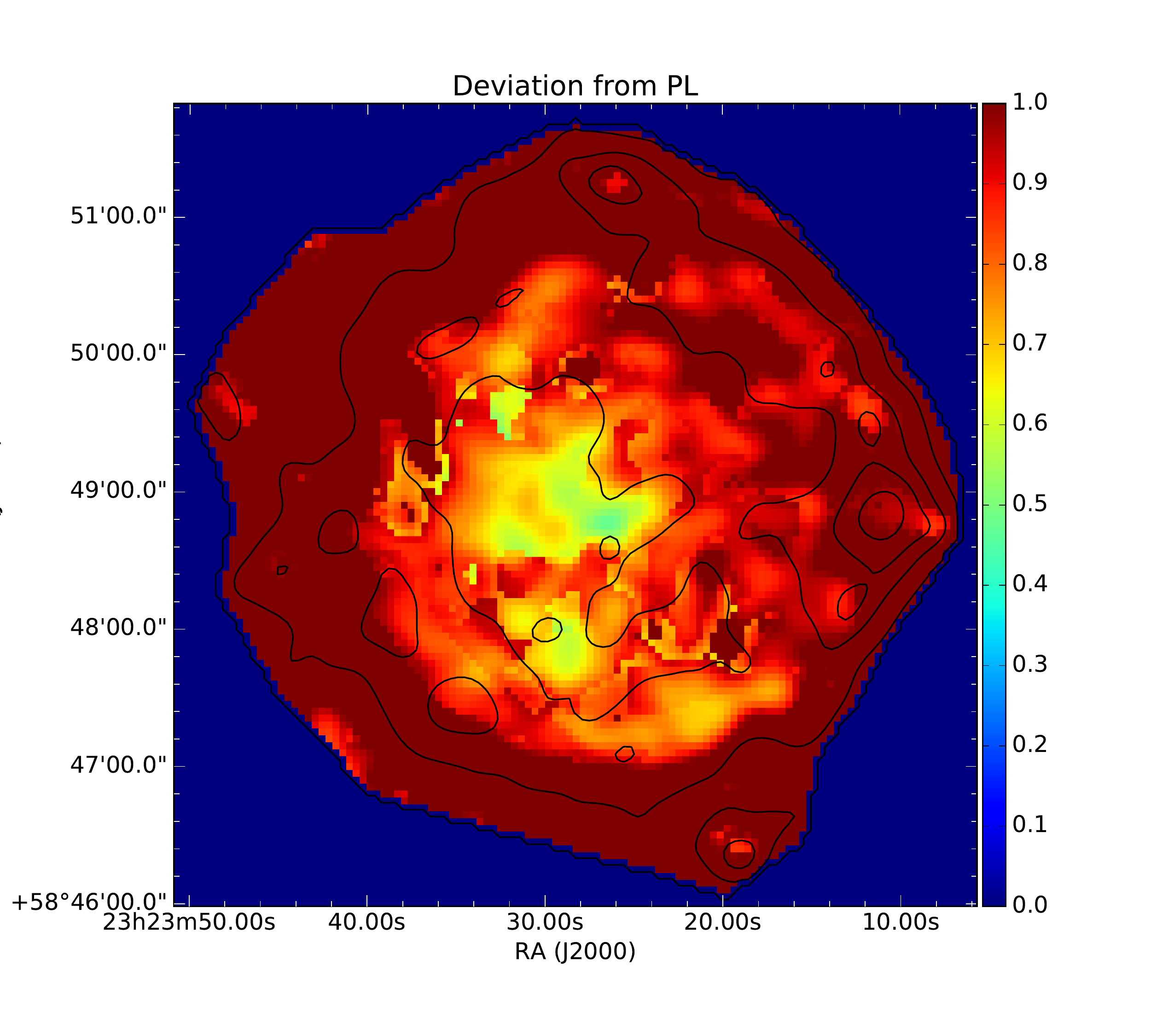}
\end{multicols}
\begin{multicols}{2}
\includegraphics[width=\columnwidth]{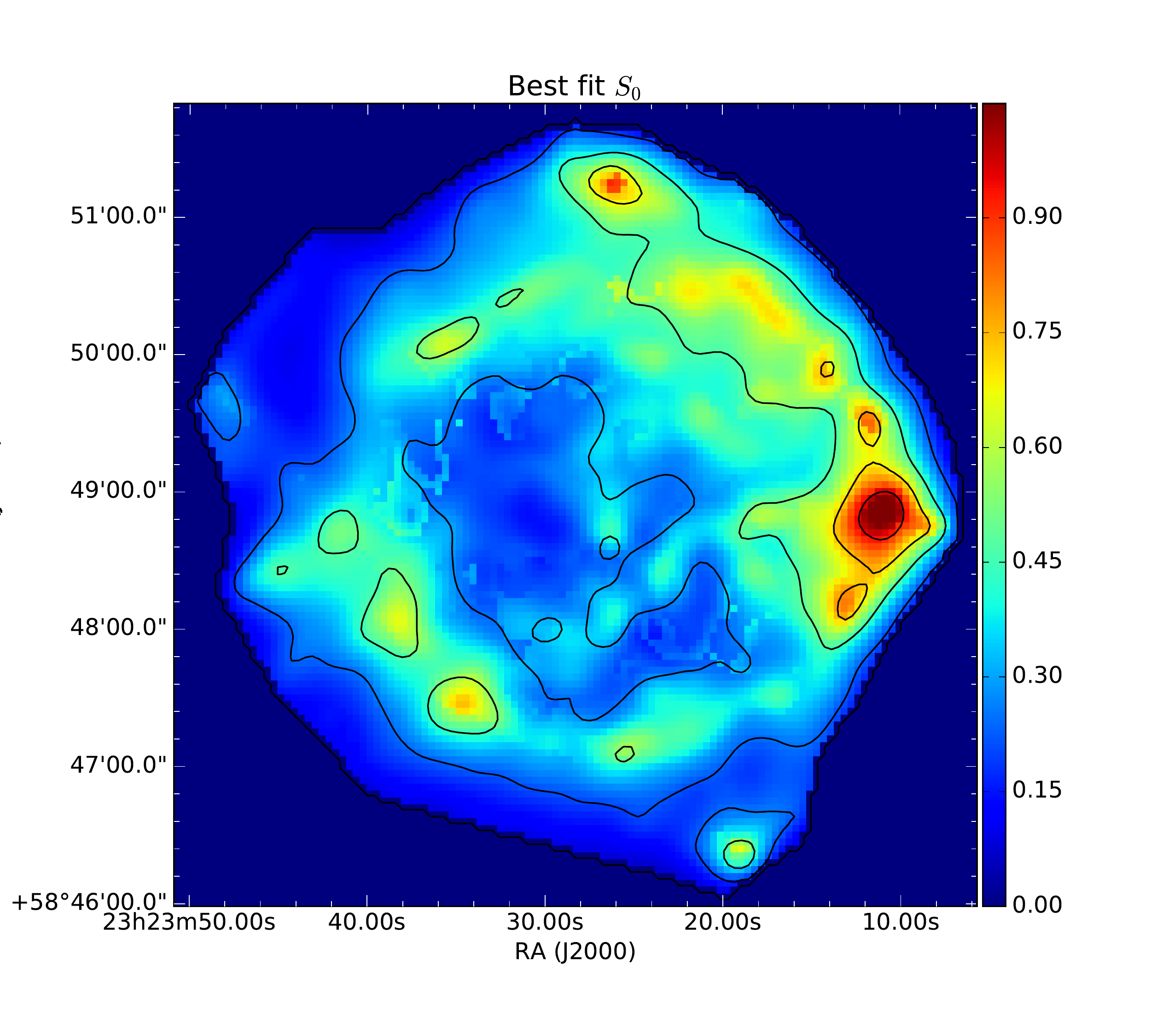}
\includegraphics[width=\columnwidth]{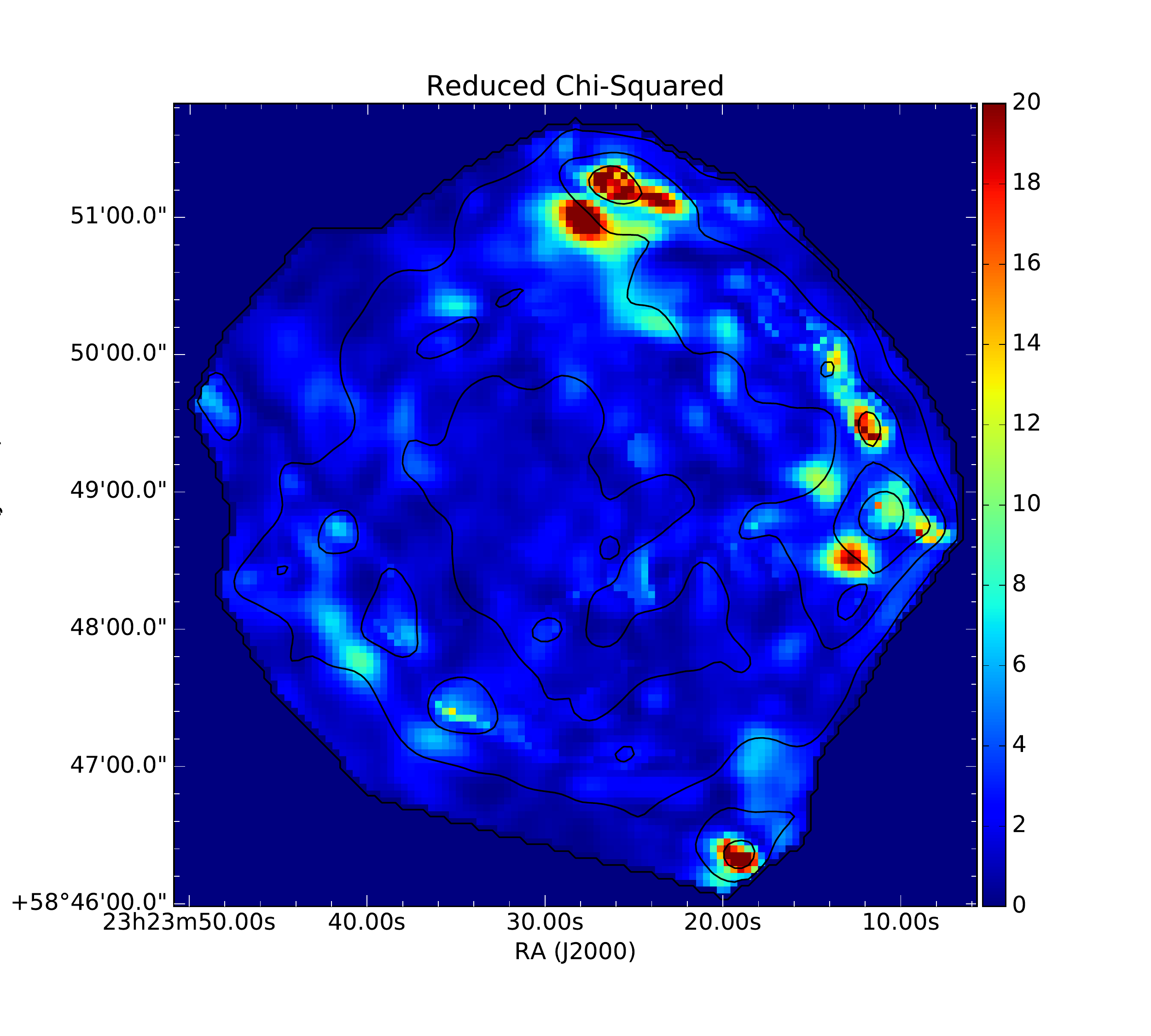}
\end{multicols}
\caption{
Results of fitting our narrow-band bootstrapped images to equation \ref{fitted_equation}. For all images the contours overlaid are at 70 MHz. \textit{Top left:} best-fit covering fraction, $f$, per pixel. No information about the location of the reverse shock is fed to the fit, but it naturally recovers $f = 1$ for regions outside the reverse-shock radius (i.e. no internal absorption). The average value of $f$ inside the reverse shock is 0.78. \textit{Top right:} deviation from power-law behaviour. This plot corresponds to $(f + (1 - f) e^{-\tau_{\nu, \mathrm{int}}})$ for our best-fit values of $f$ and $X$ (see equations \ref{fitting} and \ref{eq_x}). \textit{Bottom left:} best-fit $S_0$ per pixel. This corresponds to the flux density of Cas A at 1 GHz in jansky if no absorption were present. \textit{Bottom right:} reduced $\chi^2$ of our fit.}
\label{fit_results}
\end{figure*}

The result of this fit is shown in Fig. \ref{fit_results}. We
note that in the top row of Fig. \ref{fit_results}, no information about the location of the reverse shock is assumed a priori,
but the fit naturally recovers $f = 1$ for regions outside the reverse-shock radius (i.e. no internal absorption).
The reduced $\chi^2$ per pixel is plotted in Fig.  \ref{fit_results} (d). The higher values at the brightest knots are due to systematics that affect relatively bright point-like sources. These include both the fact that the errors are taken as constant in the image (the rms of the background pixels), and errors from the image deconvolution.

\begin{figure}
   \centering
   \includegraphics[width=8cm]{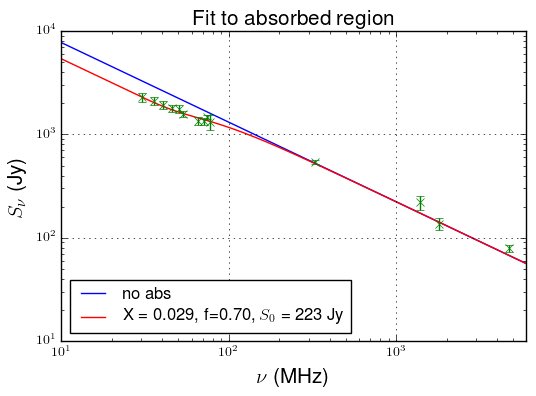}
      \caption{Fit to the absorbed region. The reduced $\chi^2$ of this fit is 1.24. 
              }
         \label{abs_reg_fit}
   \end{figure}

Figure \ref{abs_reg_fit} gives an impression of how well the data match the model. It also
illustrates the effect of
internal free-free absorption on a synchrotron spectrum.  
These data points are the sum of the flux densities, per image, of the region with high absorption in the southeast of the remnant analysed by \cite{delaney14}.

\subsection{Location of the reverse shock}

   \begin{figure*}
   \centering
   \includegraphics[width=1.5\columnwidth]{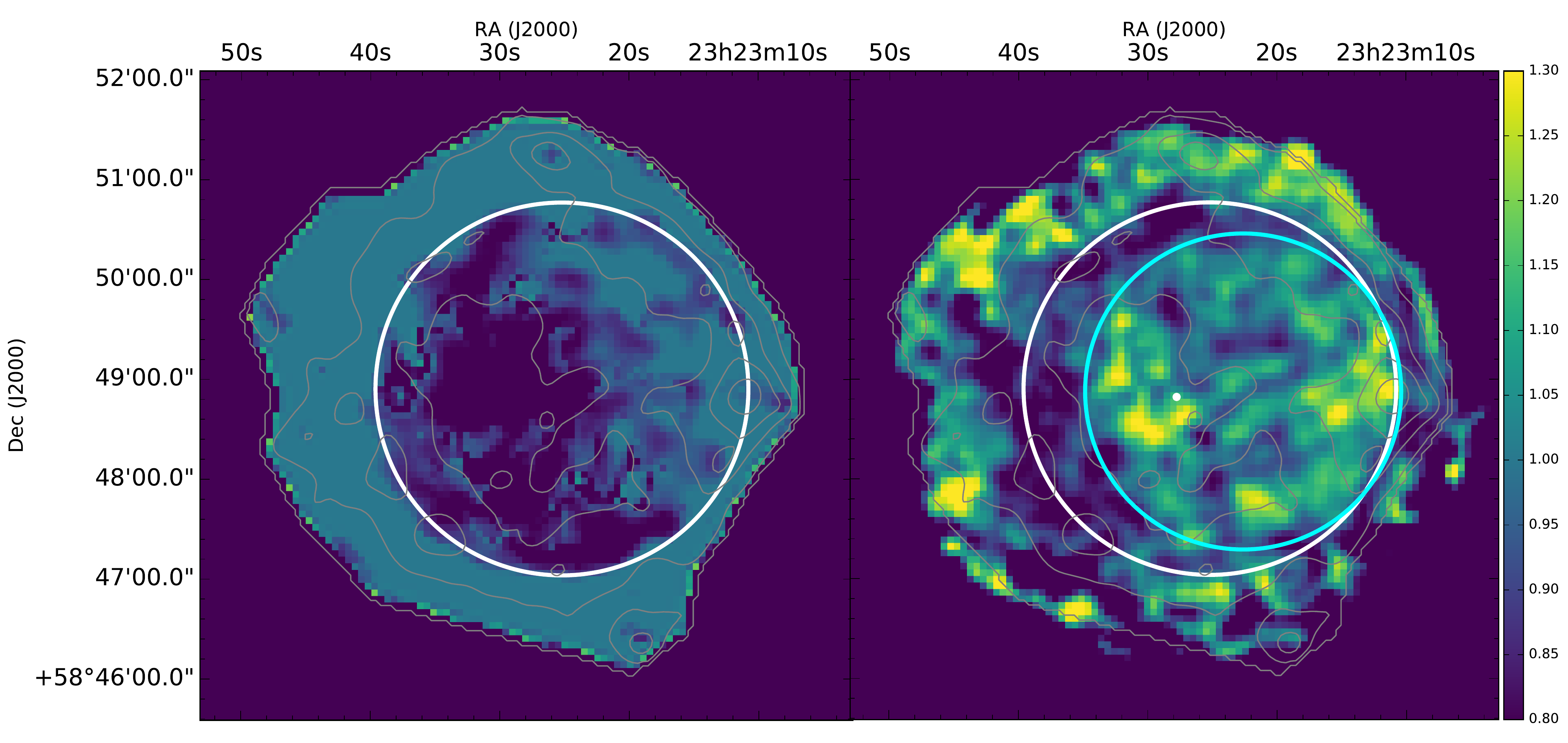}
      \caption{Comparison between the location of the reverse shock as seen in the radio and as probed by interior non-thermal X-rays. Left is Fig. \ref{fit_results} (top right). Right is a hardness ratio map with the bright parts likely indicating where non-thermal emission is dominant. The location of the reverse shock as implied from the radio map (white circle) does not match the location as seen from non-thermal X-ray filaments (cyan circle). The white dot is the expansion centre as found in \cite{thorstensen01}.
              }
         \label{xrays}
   \end{figure*}

 The top row in Fig. \ref{fit_results} shows that the internal absorption comes from a very distinct, almost circular region located roughly within the shell of Cas A.
 This region likely defines the location of the reverse shock, but it differs in several aspects from the reverse
 shock obtained from \textit{Chandra} X-ray data  \citep{helder08}, as illustrated in Fig.~\ref{xrays}.
 
Figure \ref{xrays} (right) shows a hardness ratio map made from \chandra\ ACIS-S data 5--6 keV and 3.4--3.6 keV continuum dominated bands, based
on the deep observation made in 2004 \citep{hwang04}. Here, harder regions are more likely to be synchrotron emission; the bright parts indicate where likely non-thermal emission is dominant (the forward and reverse shocks), whereas lower hardness ratios are dominated by thermal bremsstrahlung \citep[see][]{helder08}.

The images in Fig. \ref{xrays} differ in several respects. First of all, the reverse-shock radius derived from the radio is about $\sim114$\arcsec $\pm$6 \arcsec, as compared to 95\arcsec$\pm$10\arcsec\
as traced by the interior non-thermal X-ray filaments. Second, the X-ray reverse shock defines a sphere that appears to be shifted toward the western side of the SNR. 
In the western region, the location of the X-ray and radio-defined reverse-shock region coincide.

Both the radio and the X-ray data indicate a shift of the reverse shock toward the western side of the remnant. For the radio data, the approximate
centre is at 23:23:26, +58:48:54 (J2000). \cite{gotthelf01a} find the reverse shock to be centred at 23:23:25.44, +58:48:52.3 (J2000), which is in
very good agreement with our value. These should be compared to the likely explosion centre given by  \cite{thorstensen01}:
23:23:27.77, +58:48:49.4 (J2000).
The reverse shock as evidenced from the \lofar\ data is at a distance of 1.52\arcmin\
from the explosion centre at its closest point, 
and 2.2\arcmin\ at the farthest (for a distance of 3.4~kpc, 1\arcmin=1 pc).
Since the ejecta internal to the reverse shock are freely expanding, we expect them to be moving at $v_{\mathrm{ej}} = \frac{R}{t}$, which corresponds to velocities of 
4400 km\,s$^{-1}$ and 6400 km\,s$^{-1}$ for either case. 

The radio-defined reverse shock does coincide with the X-ray reverse shock in the western region, as shown in Fig. \ref{xrays}. 
The reason probably is that the X-ray defined reverse shock is based on the
presence of X-ray synchrotron emitting filaments \citep{helder08}, which requires large shock speeds
\citep[$\gtrsim 3000$~km\,s$^{-1}$,][]{zirakashvili07}. This condition is more easily met at the western side, where the reverse shock is at a larger
radius (and hence free expansion velocity) and where the reverse shock seems to move inward, increasing the velocity with which the ejecta are being
shocked. Most of the inner X-ray synchrotron emitting
filaments are indeed found in this region. This suggests that the internal radio absorption gives a more unbiased view of the location of the reverse shock,
since it does not depend on the local reverse-shock velocity.

Several other works have measured the radius of the reverse shock by tracing the inside edge of
the shocked ejecta. \cite{reed95} measured an average velocity in the optical fast-moving knots (FMKs) of 5209$\pm$90 km\,s$^{-1}$. 
The FMKs heat to optical temperatures
as they encounter the reverse shock, and so trace its rim. Their measured Doppler velocity corresponds to a reverse-shock radius of 116\arcsec$\pm$15\arcsec.
\cite{gotthelf01a} measured the reverse-shock radius by decomposing Si-band \textit{Chandra} data in radial profiles and noting a peak in emissivity at $95\pm10$\arcsec\
that, they argued, corresponds to the inner edge of the thermal X-ray shell. \cite{milisavljevic13} also conducted a Doppler study to kinematically reconstruct 
the material emitting in the optical. They find that the reverse shock is located at a velocity of 4820 km\,s$^{-1}$, which corresponds to $106\pm14$\arcsec. These values agree within the error bar
with each other, as well as with our absorption-derived one.

\subsection{Is there evidence for synchrotron self-absorption?}

\cite{atoyan00} proposed that Cas A might have dense, bright knots with a high magnetic field ($\sim 1.5$ mG) within a diffuse region of low magnetic field. These knots would begin to self-absorb at the frequencies where the brightness temperature  approaches the effective electron temperature $T_\mathrm{e}$. 

Synchrotron electrons have effective temperatures:
\begin{equation}
T_\mathrm{e} = \frac{1}{3k} \sqrt{\frac{\nu_\mathrm{c}}{1.8 \times
  10^{18} B}},
  \label{temp}
\end{equation}
where $\nu_c$ is the critical frequency and $E = 3k T_\mathrm{e}$ for a relativistic gas. For a blackbody in the Raleigh-Jeans approximation,
\begin{equation}
I_\nu = \frac{2kT\nu^2}{c^2}.
\end{equation}

Since $S_\nu = I_\nu \Omega \approx I_\nu \theta^2$, and $I_\nu$ is at most as large as the emission from a blackbody, substituting for the temperature value in equation \ref{temp}, we arrive at\begin{equation}
\frac{S_\nu}{\theta^2} \leq \frac{2}{3} \frac{1}{c^2}
\frac{\nu^{5/2} B^{-1/2}}{\sqrt{1.8 \times
  10^{18}}} ,
\end{equation}
with $\nu$ in hertz, $\theta$ in radians, and $B$ in gauss. 

It is possible to use this relation to determine at which frequency $\nu$ we would expect the synchrotron spectrum of a source of angular size $\theta$ and magnetic field $B$ to peak (i.e. roughly begin to be affected by self-absorption). We used  the flux densities in Fig.  \ref{fit_results} (c) to calculate the synchrotron self-absorption frequency for each pixel if all the remnant were to have the (high) magnetic field of 1.5 mG proposed by \cite{atoyan00}. We find that the break frequencies are only as high as $\sim 8$ MHz for the brightest knots and $\sim 4$ MHz for the more diffuse regions of the remnant. Features more compact than our pixel size $\theta=3\arcsec$ could self-absorb at \lofar \, frequencies, but are not resolved.

\section{Interpretation of internal absorption}

\subsection{Internal mass}

A measured value of internal free-free absorption alongside assumptions about the source geometry and physical conditions allows us to constrain two physical parameters: the internal electron density, and the mass.

From the best fit to our images we obtain a value for a combination of the emission measure $EM$, the temperature $T$, and the average number of charges of the ions $Z$. As noted before, $EM =  \int_{0}^{s'} n_\mathrm{e}^2 ds'$, so $EM$ is the parameter that we need in order to obtain a mass estimate
of the unshocked ejecta. This requires us to fix a value of $T$ and $Z$. Moreover, solving for $n_\mathrm{e}$ requires assumptions about the geometry of the ejecta. If $n_\mathrm{e}$ is constant inside the reverse shock, then $EM = n_\mathrm{e}^2 l$, where $l$ is a thickness element. 

The total mass of unshocked ejecta is its density times its volume, $M_{\mathrm{unsh}} = \rho V$. The ions are the main contributors to the mass, and the density of ions is their number density $n_\mathrm{i} = \frac{n_\mathrm{e}}{Z}$ times their mass, $A m_\mathrm{p}$, where $A$ is an average mass number, $m_\mathrm{p}$ is the mass of the proton, and $Z$ is the ionisation state (and not the atomic number).  Hence we obtain $\rho = A m_p  \frac{n_\mathrm{e}}{Z}$. 

The volume $V$ associated with a given pixel is related to the thickness element $l$ in the following way: $V = S l$, where $S$ is the projected surface area (in the case of our image, the $3 \arcsec \times 3 \arcsec$ pixel). The total mass in the unshocked ejecta in the case of constant density for each given pixel is
\begin{equation}
M =  A S l^{1/2} m_\mathrm{p} \frac{1}{Z} \sqrt{EM}.
\label{eq_mass}
\end{equation}

The measured value of $EM$ depends weakly on $Z$ and is quite sensitive to $T$. In addition, given the dependency of the unshocked ejecta mass on surface area $S$ and length $l$, any estimate critically depends on assumptions about its geometry. 
This is why the images in Fig. \ref{fit_results}  are more fundamental, as they correspond to the directly measured parameters. 
No assumptions about the shape, composition, ionisation state, or temperature enter the fitting for $X$.

\subsection{Emission measure}\label{sec_em}

In order to convert our best-fit values of $X$ into an emission measure map, we take the following steps:
   \begin{enumerate}
      \item We take only values internal to the reverse shock, since these are the values that are relevant to internal free-free absorption.
      \item We mask the values that correspond to $f < 0.1$ and $f > 0.9$. These extreme values might be due to pixel-scale artefacts in the images; moreover, for values of $f \sim 1,$ the value of $X$ is degenerate (see equation \ref{fitting}). 
      \item We assume that in the plasma internal to the reverse shock $T = 100$~K, and $Z=3$. These values are proposed in \cite{eriksen09}.
   \end{enumerate}

\begin{figure}
   \centering
   \includegraphics[width=\columnwidth]{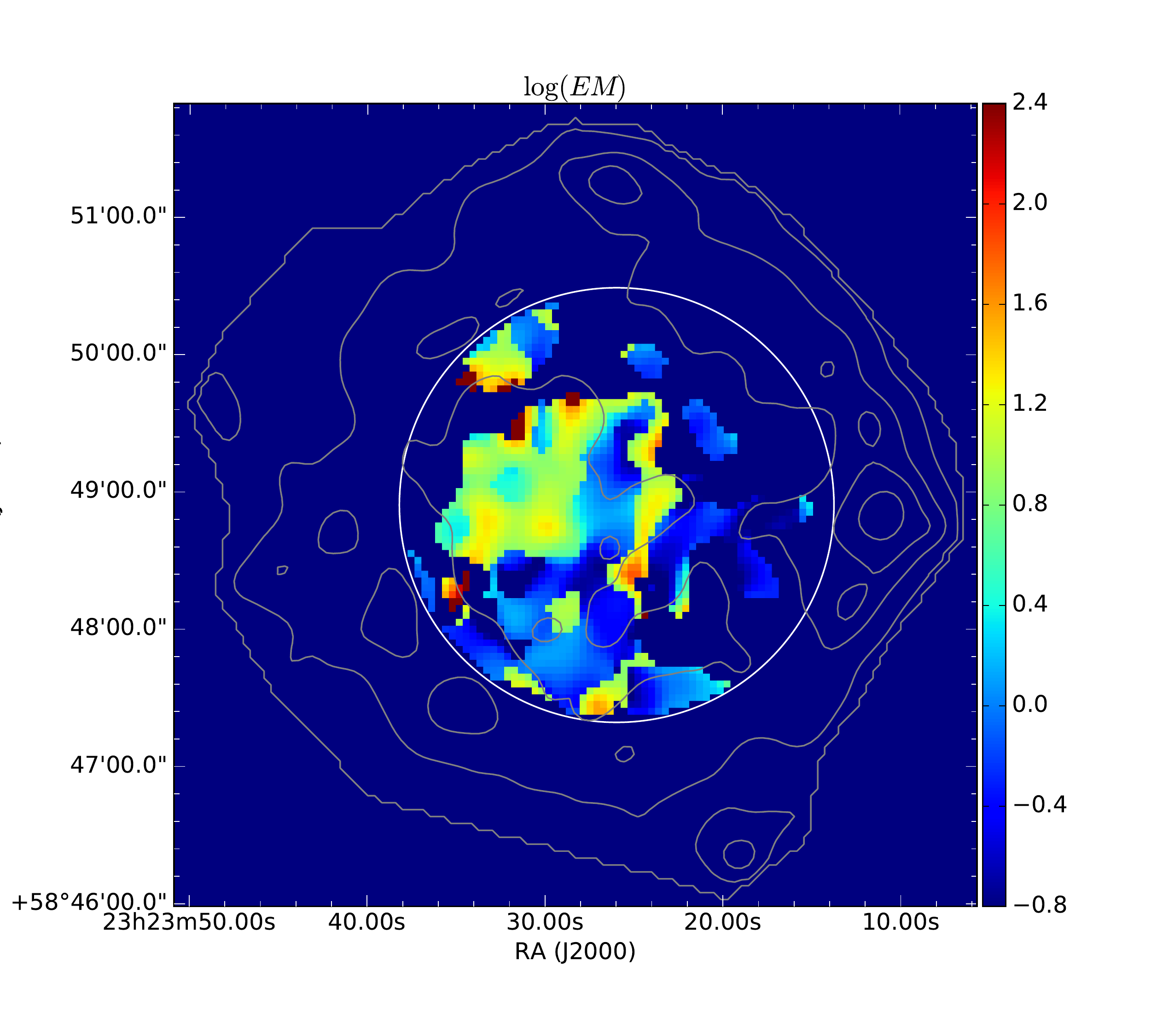}
      \caption{$log_{10}$ of the emission measure value per pixel. When used to calculate the mass, the blanked pixels are approximated by the average of the remaining $EM$  values, $EM = 37.4~\rm{pc} \, \rm{cm}^{-6}$.
              }
         \label{log_EM}
   \end{figure}

Using equation \ref{eq_x} and solving for $EM$, we obtain Fig. \ref{log_EM}. In the case of the fit to the absorbed region shown in Fig. \ref{abs_reg_fit}, the best-fit $X$ for the same temperature and ionisation conditions implies $EM = 7.1 \, \rm{pc} \, \rm{cm}^{-6}$.

A caveat with this analysis is that within the reverse-shock radius, we blanked out a significant fraction of the pixels, since our best fit
indicated that for most of these pixels $f>0.9$. This likely means that in these regions most of the radio emission is dominated by the front side of the shell.
We did find that the fitted values for emission measure in these pixels were much higher
than those shown in Fig.~\ref{log_EM}. The degeneracy between $f$ and $EM$ implies that we cannot trust these high values, but 
these blanked-out regions should still contribute to the overall mass budget of unshocked ejecta even if we cannot access them because of the geometry of the shell.

In order to account for the mass associated with these blanked-out pixels, we assumed that the $EM$ in these pixels was equal to the average $EM$
from the selected pixels: $EM = 37.4 \, \rm{pc} \, \rm{cm}^{-6}$. Since we  blanked out 40\% of the pixels, this does imply a systematic error of a similar
order in the mass estimate given below.

\subsection{Mass estimate}

As mentioned previously, any mass estimate critically depends on the assumed geometry of the unshocked ejecta. These are clumpy, asymmetric, and notoriously difficult to trace. \cite{isensee10} pointed out that the O, Si, and S ejecta can form both sheet-like structures and filaments from infrared observations. \cite{milisavljevic15} expanded on this view by proposing that Cas A has a cavity-filled interior with a \lq Swiss cheese' structure.

We do not know what the shape of the unshocked ejecta is behind every pixel in our image. The work of \cite{delaney14}  uses a geometry where the unshocked ejecta is confined to two sheets interior to the reverse shock in order to obtain a mass estimate from an optical depth measurement. They take these sheets to be 0.16 pc thick and have a total volume of 1.1 pc$^3$.

Using these same parameters, our estimate of the mass in the unshocked ejecta is 

\begin{equation}
\begin{split}
M = & 2.95 \pm ^{0.41} _{0.48} M_{\odot}\, \left(\frac{A}{16}\right) \left(\frac{l}{0.16 \rm{pc}}\right)^{1/2} \left(\frac{Z}{3}\right)^{-3/2} \left(\frac{T}{100~\mathrm{K}}\right)^{3/4} \\
& \times \sqrt{\frac{g_{\mathrm{ff}}(T=100 \, \mathrm{K},Z=3)}{g_{\mathrm{ff}}(T,Z)}}.
\end{split}
\label{eqn_mass}
\end{equation}
The errors here are the statistical errors of the fit. The systematic error due to our blanking of some pixels is of order 40\%.

This estimate is puzzling if we consider that Cas A is thought to have a progenitor mass before the explosion of 4 -- 6 \msun \, \cite[]{young06}, and that most of the ejecta is presumed to have already encountered the reverse shock. We discuss this issue further in this section, but 
note here that the estimate is sensitive to the geometry ($l$) and composition ($Z$, $A$) of the unshocked ejecta. 

For the same parameters, we estimate the electron density in the unshocked ejecta $n_\mathrm{e} = \sqrt{\frac{EM}{l}}$ to be\begin{equation}
\begin{split}
n_e =& 18.68 \pm ^{2.62} _{3.05} \mathrm{cm^{-3}} \left(\frac{0.16 \mathrm{pc}}{l}\right)^{1/2}  \left(\frac{Z}{3}\right)^{-1/2} 
\left(\frac{T}{100~\mathrm{K}}\right)^{3/4} \\
& \times \sqrt{\frac{g_\mathrm{{ff}}(T=100 \, \mathrm{K},Z=3)}{g_{\mathrm{ff}}(T,Z)}}.
\end{split}
\label{eq_n_e}
\end{equation}

If we consider only the area that \cite{delaney14} studied, using the same parameters as above, our mass estimate is $1.15$ \msun, and $n_\mathrm{e} = 6.65 \,\rm{cm}^{-3}$. For $T = 300$ K, and $Z = 2.5$ (the parameters employed in that work),  the mass estimate for this region is $2.50$ \msun, and $n_\mathrm{e} = 14.37 \,\rm{cm}^{-3}$.

\subsection{Comparisons to earlier results}
\label{section:others}

\cite{delaney14} estimated a mass of 0.39 \msun\ in unshocked ejecta, but in their derivation of the unshocked mass from the measured absorption,
they confused ion charge, atomic number (both often denoted by the same symbol $Z$) and atomic mass number, as detailed in Footnote 1.
Their measured quantity is the optical depth at 70 MHz,  $\tau_{70 \mathrm{MHz}} = 0.51$. Our best-fit value of $X$ (Eq.~\ref{eq_x}) for the same region they analyse implies an optical depth of $\tau_{70 \mathrm{MHz}} = 0.97$, with the additional consideration that only 30\% of the flux density in that region comes from the backside of the shell and is subject to being absorbed. 

For their measured optical depth, using ion charge $Z=3$ (as opposed to $Z = 8.34$), the derived electron density is $n_\mathrm{e} = 12.9$ cm$^{-3}$ (as opposed to the 4.23 cm$^{-3}$ that they quote). With the geometry described above, and using  $\rho = A m_\mathrm{p}  \frac{n_\mathrm{e}}{Z}$ (that is, multiplying by the mass number and not the atomic number as they do), their mass estimate is in fact 1.86 \msun. 
Moreover, they extrapolated the optical depth of a limited region to the whole area inside the reverse-shock radius,
although their Fig.~7 and  Fig.~ \ref{log_EM} in this paper both
indicate that there are substantial variations. We have a similar limitation
concerning the blanked-out regions with the reverse shock (see Sect.~\ref{sec_em}).

Our results are therefore different by around a factor of 3/2 from those in \cite{delaney14}, but we note that our measurements are based
on fitting per pixel, including the parameter $f$, using a broader frequency coverage, and including lower frequencies for which the absorption effects are more pronounced.
Both our absorption values and those of \citet{delaney14} imply masses that are relatively high, as discussed  further below.  Eq.~\ref{eqn_mass} shows that
the mass estimate depends strongly on the temperature, as well as clumping, mean ion charge, and composition. In the next section we discuss the effects
of these factors on the mass in unshocked ejecta.

\section{Discussion}

Our derived value of mass in the unshocked ejecta from our measured low-frequency absorption for a gas temperature of 100 K,
 an ionisation state of 3, and a geometry where the ejecta are concentrated in relatively thin and dense sheets is of the order of 3 \msun. This value is at odds with much of the conventional wisdom on Cas A, but is not that much higher than the value estimated in the low-frequency absorption work of \cite{delaney14}, see  Sect.~\ref{section:others}. In this section, we attempt to reconcile our low-frequency absorption measure with constraints from other observed and modelled features of Cas A.

\subsection{Census of mass in Cas A}

The progenitor of Cas A is thought to be a 15--25 \msun\ main-sequence mass star that lost its hydrogen envelope to a binary interaction \cite[]{chevalier03,young06}. It is difficult to determine the mass of the progenitor immediately before the explosion, although the star must have lost most of its initial mass to explode as a Type IIb \cite[]{krause08}. 

Approximately 2 \msun \, of the progenitor mass transfer into the compact object, which is thought to be a neutron star \cite[]{chakrabarty01}. The shock-heated ejecta accounts for 2--4 \msun\ of material. This value is obtained from X-ray spectral line fitting combined with emission models \cite[]{vink96,willingale02}. \cite{young06} noted that if this is a complete census of the Cas A mass (this ignores any mass in the unshocked ejecta, and also any mass in dust), combined with constraints from nitrogen-rich high-velocity ejecta, and $^{44}$Ti and $^{56}$Ni abundances, then the total mass at core collapse would have been  4--6 \msun. 
\cite{lee14} proposed 5 \msun\ before explosion from an X-ray study of the red supergiant wind.
These values  for the progenitor mass immediately before explosion are used in a number of models that reproduce the observed X-ray and dynamical properties of Cas A. For instance, the observed average expansion rate and shock velocities can be well reproduced by models with an ejecta mass of $\sim 4$ \msun\ \citep{orlando16}.  

Models for the interaction of the remnant with a circumstellar wind medium indicate that the reverse shock in Cas A has already interacted with a significant fraction of the ejecta \cite[]{chevalier03}. \cite{laming03} applied their models directly to \textit{Chandra} X-ray spectra and also inferred that there is very little unshocked ejecta remaining (no more than 0.3 \msun). 

On the other hand, \cite{delooze17} find a surprisingly high SN dust mass between 0.4 -- 0.6 \msun, which is at odds with \citet{laming03}. 
Given the uncertainties in mass estimates from both observational and theoretical considerations, a total
unshocked ejecta mass of $\sim 3$ \msun \, is high, but not impossible. 
Here we discuss several properties that may affect the mass estimate from the radio absorption measurements.

\subsection{Effect of clumping on the mass estimate}

The most significant of these effects has to do with the geometry of the ejecta. 
The infrared Doppler shift study of \cite{isensee10} and the ground-based sulphur observations of \cite{milisavljevic15} provide clear evidence that the unshocked ejecta is irregular.
One way to avoid lowering the mass estimate while maintaining the measured absorption values is to consider  the effect of clumping.

We assume that the unshocked ejecta consist of 
two zones: one zone composed of $N$ dense clumps, and a diffuse, low-density region. It is possible that the dense region contributes a small amount toward the total mass, but is responsible for most of the absorption. We describe the density contrast by the parameter $x\equiv n_{\mathrm{clump}}/n_{\mathrm{diff}}$, and denote the
typical clump radius by the symbol $a$. All the unshocked ejecta is within the radius of the reverse shock $R_\mathrm{rev}$.

The optical depth in the diffuse region is given by $\tau_{\mathrm{diff}} \propto n_{\mathrm{diff}}^2R_\mathrm{rev}$, and for each clump $\tau_{\mathrm{clump}} \sim (x n_{\mathrm{diff}})^2 a$. 
For clumps to be responsible for most of the absorption, we require $x^2 a \gg R_\mathrm{rev}$. 
We call $p$ the surface area filling factor for the clumps, $0 < p < 1$. If the clumps are compact, there is likely not more than one clump in a single line of sight, so that 
$ N \pi a^2 \approx p \pi R_\mathrm{rev}^2$, and therefore\begin{equation}
a \approx \sqrt{\frac{R_\mathrm{rev}^2 p}{N}} \gg \frac{R_\mathrm{rev}}{x^2}.
\label{eq_a}
\end{equation}
The total mass is (see Eq. \ref{eq_mass})
\begin{equation}
\begin{split}
M & = M_{\mathrm{diff}} + M_{\mathrm{clump}} = \frac{4 \pi}{3} A m_\mathrm{p} \frac{n_{\mathrm{diff}}}{Z} \left( R_\mathrm{rev}^3 + N a^3 x \right) \\
 & = \frac{4 \pi}{3} A m_\mathrm{p} \frac{n_{\mathrm{diff}}}{Z}  R_\mathrm{rev}^3 \left(1 + \frac{p^{3/2}}{\sqrt{N}} x \right).
\end{split}
\end{equation}
For the diffuse component to dominate the mass estimate, i.e. $M \approx M_{\mathrm{diff}}$, we need $ p^{3/2}x \ll \sqrt{N}$, while $\sqrt{\frac{p}{N}} x^2 \gg 1$ (Eq. \ref{eq_a}). 

In order to lower our mass estimate by a factor of 100, we require $n_\mathrm{diff} = 0.1 \, \rm{cm}^{-3}$. In this case, the material in dense clumps that is responsible for the absorption should have $EM = n_\mathrm{e}^2 l$ such that $37.4 \, \rm{pc \, cm}^{-6}$ $= ( 0.1 \, x \, \rm{cm}^{-3})^2$ $l$. With $l \sim a$, this implies $x^2 a \sim$ 4000 pc . Using Eq. \ref{eq_a} and $R_\mathrm{rev} = 1.58$ pc, we arrive at $x^2 \sim 2500 \sqrt{\frac{N}{p}}$, which combined with the condition that $\sqrt{\frac{p}{N}} x \ll 1$, gives $x \gg 2500$. The other condition, $ p^{3/2}x \ll \sqrt{N}$, implies $2500 p \ll x$. $p$ can be at most 1, so the second condition is fulfilled whenever the first one is.

The required ratio of the densities of the clumped and diffuse media gives knot densities $n_{\mathrm{clump}} \gg 250 \, \rm{cm}^{-3}$. This is in line with the densities in fast-moving shocked optical knots measured in \cite{fesen01a}. 

Accounting for clumping can significantly lower the estimated mass, although it would be contrived to match our observations with the models that predict almost no mass in the unshocked ejecta.

\subsection{Can the unshocked ejecta be colder than 100 K?}

In addition to the effect of clumping,  Eq.~\ref{eqn_mass} shows that the mass estimate is  also very sensitive to the temperature $T$ of the unshocked gas and its ionisation state $Z$. It could be lowered if the temperature of the unshocked gas were lower than the 100 K we assume.

The temperature of the plasma interior to the reverse shock was estimated in \cite{eriksen09} from \spitzer \, observations. They argued that strong [O IV] [Si II] but weak or absent [S IV] and [Ar III] imply $T \sim$ 100 -- 500 K, but it is not clear how they estimated the temperature of the unshocked gas from the
line ratios. 
In the case of a rapidly expanding gas, one needs to be careful about temperature measurements from line ratios, as the
ionisation balance can be out of equilibrium (especially since the ionisation is likely dominated by photo-ionisation).
Under equilibrium conditions, the recombination timescale and ionisation timescale are equal. These conditions fail, however,
if the recombination timescales are longer than the age of the SNR.

There is some reason to believe that the temperature inside the reverse shock might be lower than 100 K. 
\cite{delooze17} measured the temperature of the dust components inside the reverse shock to be approximately 35 K (and we note that in the ISM, the gas is normally colder than the dust, although the reverse is true in some regions, such as the galactic centre). 
Moreover, if we set the energy density in the infrared field equal to that of a blackbody and find an associated temperature, this is of the order of 10 K: 

The SED of Cas A is dominated by the contribution of the infrared, and we take this to be an approximation of the total. The energy density is the luminosity in the infrared, divided by the volume internal to the reverse shock, multiplied by the average amount of time a photon spends in the inside of the reverse shock, that is, 
\begin{equation}
u_\mathrm{R} =F_\mathrm{IR} 4 \pi d^2 \frac{1}{\frac{4}{3} \pi R_\mathrm{rev}^3} \frac{\frac{4}{\pi} R_\mathrm{rev}}{c} = \frac{4 \sigma}{c} T^4.
\end{equation}
When we use $F_{\mathrm{IR}} = 2.7 \times 10^{-8}$ erg cm$^{-2}$ s$^{-1}$ \cite[]{arendt89} and $d = 3.4$ kpc, the distance to Cas A \cite[]{reed95x} is $T \sim 10$~K. 

A full non-equilibrium photo-ionisation treatment is beyond the scope of this paper. Here we limit ourselves to pointing out that for the observed ionised oxygen species up to O IV,  the recombination timescales are longer than the age of the remnant, even for a temperature as cold as 10 K. 

\begin{figure}
   \centering
   \includegraphics[width=9cm]{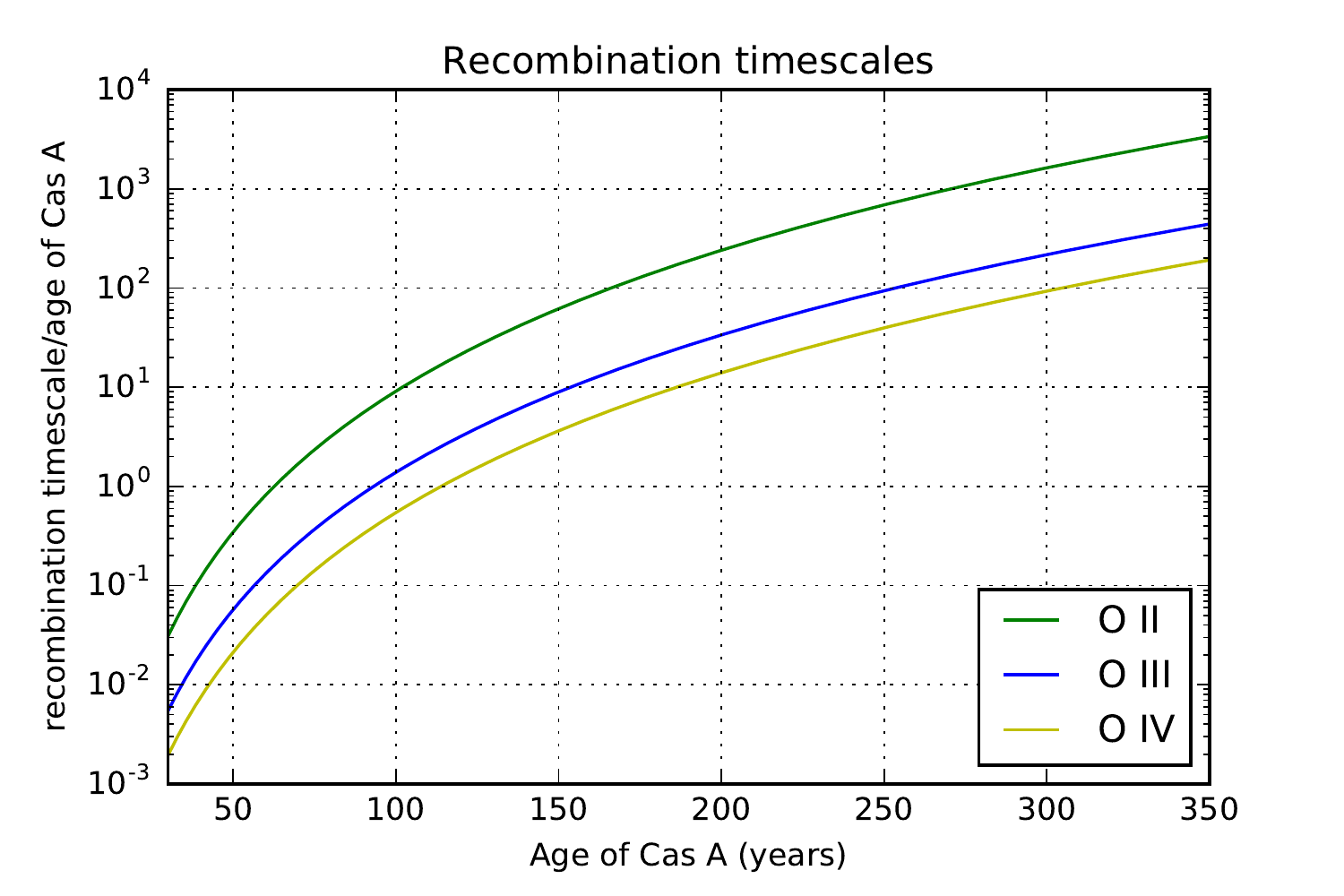}
      \caption{Recombination timescales as a fraction of the age of the remnant $\frac{1}{n_\mathrm{e} \beta_{\mathrm{rad}}}\frac{1}{t}$ for a gas that is expanding adiabatically and that is normalised so that in 2015, $T=10$ K and $n_e = 10 \, \mathrm{cm}^{-3}$.
              }
         \label{recon_timescales}
   \end{figure}

The radiative recombination coefficients $\beta_{\mathrm{rad}}$ for oxygen ions are given in Table 7.3 of \cite{tielens05}. 
In Fig. \ref{recon_timescales} we plot the recombination timescales as a fraction of the age of the remnant $\frac{1}{n_\mathrm{e} \beta_{\mathrm{rad}}}\frac{1}{t}$ for a gas that is expanding adiabatically and that is normalised so that in 2015 ($t = 343$ yr), $T=10$ K and $n_e = 10 \, \mathrm{cm}^{-3}$. 
The recombination timescales become longer than the age of the remnant within the first 150 years for all three species. If we include the effect of clumping, the majority of the mass is in the lower density region, and so the recombination timescales become longer than the age of the remnant even at earlier times.
This means that once an atom is ionised, it stays ionised, even though the temperature of the gas is cold. 

\subsection{Energy requirements}

Our estimate of $n_\mathrm{e} \sim 10 \, \rm{cm}^{-3}$ implies that several solar masses of material internal to the reverse shock have to be ionised, which requires a significant energy input. With the volume we used for our absorption calculations $V = \pi R_{\mathrm{\mathrm{rev}}}^2 l$, where $l=0.16$ pc, we have a total number of electrons inside the reverse shock of $3.07 \times 10^{56}$. If we assume that all of these are oxygen atoms, it takes 13.6 eV to ionise each of them a first time, 35.1 eV for a second time, and 54.9 eV for a third ionisation.  Not all oxygen atoms are ionised to the higher states, but these quantities correspond to roughly $10^{46}$ erg over the lifetime of Cas A, or $10^{36}$ erg s$^{-1}$ on average.

 For an X-ray flux of $9.74 \times 10^{-9}$ erg s$^{-1}$ cm$^{-2}$ \cite[]{seward90} and a distance of 3.4 kpc \cite[]{reed95}, the X-ray luminosity of Cas A is $1.35 \times 10^{37}$ erg s$^{-1}$. Considering transparency effects and the short recombination timescales at early times (see Fig. \ref{recon_timescales}), it is unlikely that the X-ray photons alone could maintain this amount of material ionised. 
 The high-ionisation state of the unshocked ejecta 
 therefore requires an additional source of ionisation, which could be the UV emission from the shell of
 Cas A. This emission component is difficult to measure because
of the high extinction toward Cas A.


\section{Effects of internal absorption on the secular decline of the radio flux of Cas A}

   \begin{figure}
   \centering
   \includegraphics[width=9cm]{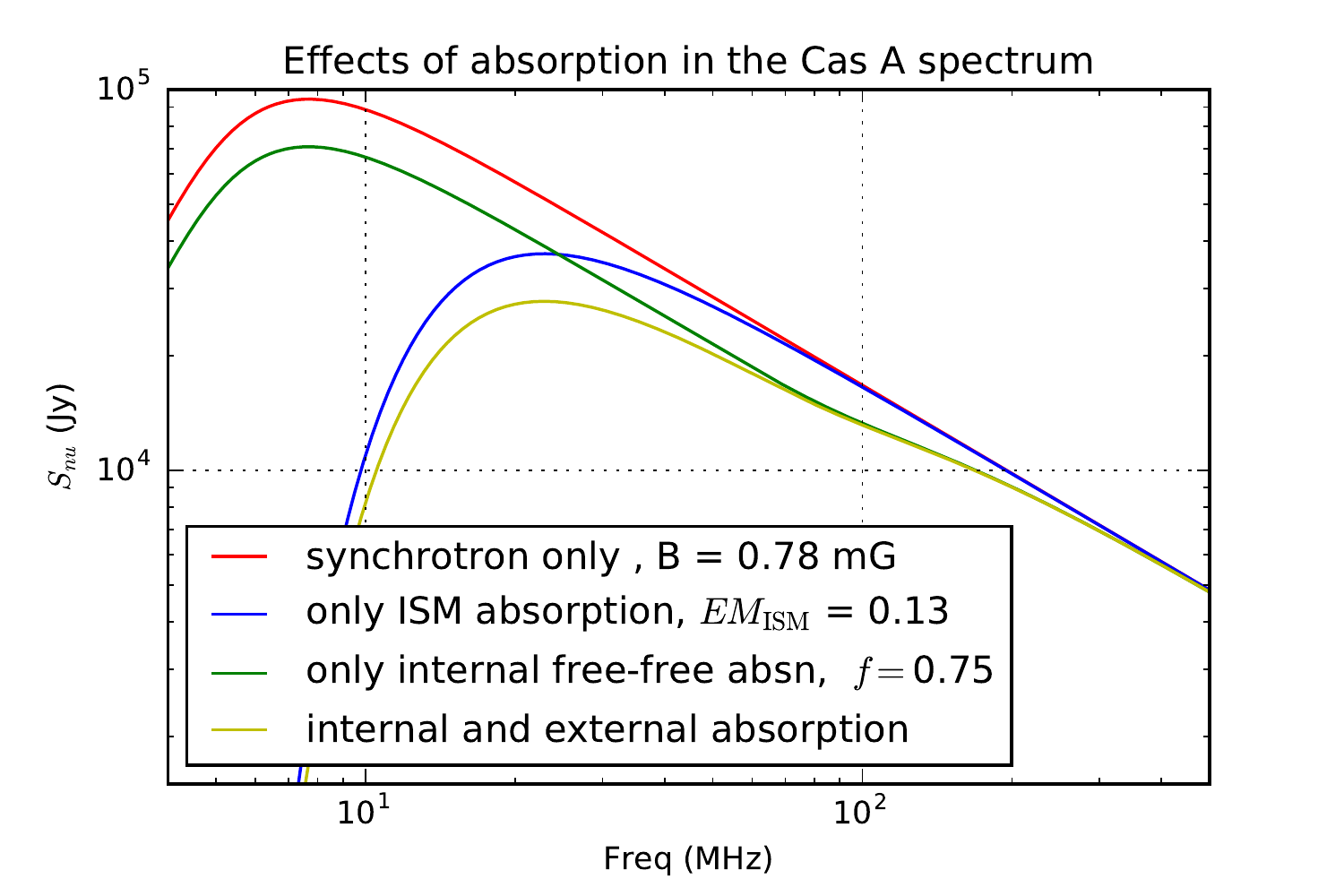}
      \caption{Effect of different forms of absorption on the integrated spectrum of Cas A for a magnetic field value of 0.78 mG. At $\sim 5$ MHz, the synchrotron emission begins to self-absorb and has a slope of $\nu^{5/2}$. The blue line shows the spectral shape that the radio source with the synchrotron spectrum shown by the red line would have if it encountered ISM absorption along the line of sight, and the green is the shape it would have if 25\% of its synchrotron shell were subject to internal free-free absorption. The yellow line is a combination of both effects.
              }
         \label{absn_effect}
   \end{figure}

Given its status as one of the brightest radio sources in the sky, the radio spectrum of Cas A has been analysed extensively. 
In this section we model the effect that internal absorption has on the integrated radio spectrum of Cas A and on its secular decline, giving a physically plausible model.

\subsection{Effect of absorption on the synchrotron spectrum}

The full expression of synchrotron emission \cite[]{longair11} is \begin{equation}
S_{\nu, \mathrm{synch}} = I_{\nu, \mathrm{synch}} \Omega = \Omega \frac{J(\nu)}{4 \pi \chi_\nu} \big( 1 - e^{- \chi_\nu l} \big),
\label{rad_transfer}
\end{equation}
where $\Omega$ is the angular size subtended by the source, $l$ is the thickness of the synchrotron emitting slab, and $J(\nu)$ and $\chi(\nu)$ are the synchrotron emission and absorption coefficients. 
 
The synchrotron flux density depends on the magnetic field strength $B$ and on the number of electrons through $\kappa$, where $N(E) = \kappa E^{-p}$ and $N(E)$ is the electron energy distribution. In principle, it is not possible to tell the two contributions apart. If we assume there is no absorption at 1 GHz, we can ignore the absorption part of Eq.  \ref{rad_transfer}, and set 
\begin{equation}
S_{1 \mathrm{GHz}} = \frac{L_{\mathrm{1 GHz}}}{4 \pi d^2} = \frac{J_{ \mathrm{1 GHz}} V}{4 \pi d^2} = 2720 \, \rm{Jy,}
\end{equation}
and in this way, we can obtain a relation between $\kappa$ and $B$:
\begin{equation}
\kappa(B) = \frac{L_{ \mathrm{1 GHz}}}{A(\alpha) V B^{1+ \alpha} (10^9)^\alpha}.
\end{equation}
This means that Eq.~\ref{rad_transfer}  can be rewritten in terms of $B$, the luminosity at 1~GHz, and the volume,
which we set to be the shell formed by the forward and reverse shocks, $V = \frac{4 \pi}{3} (R_{\mathrm{forw}}^3 - R_{\mathrm{rev}}^3)$. Hence, the only free parameter left to fit is the magnetic field  $B$,
whose strength determines the location of the low-frequency turnover.

As pointed out earlier, the unshocked ejecta only absorb a fraction of the synchrotron emitting shell. The contribution of the unshocked component to the total radio spectrum reads
\begin{equation}
S_{\mathrm{int \,abs}} = S_{\mathrm{front}} + S_\mathrm{{back}} e^{-\tau_{\mathrm{int}}} =  S_{\mathrm{synch}} (f+ (1-f)e^{-\tau_{\mathrm{int}}}).
\label{internal}
\end{equation}

The most natural component for the low-frequency radio absorption of a galactic source is free-free absorption from ionised gas in the ISM between us and the source. Our line of sight to Cas A intersects a large molecular cloud complex that includes clouds both in our local Orion arm and in the Perseus arm of Cas
A, as evidenced by absorption lines of H I and a number of other molecules that trace cold gas, such as carbon monoxide \cite[]{ungerechts00}, formaldehyde \cite[]{batrla83}, and amonia \cite[]{batrla84}. \cite{oonk17} and \cite{salas17} detected a series of carbon radio recombination lines (CRRL) in the Perseus arm towards Cas A, and were able to model a number of physical parameters in this gas. Their results are summarised in Table \ref{table:1}. These emission measure values are only lower limits, as they do not include the Orion arm clouds and only measure the cool
gas with high column density. 

\begin{table}        
\caption{Absorbing gas along the line of sight to Cas A as in \cite{oonk17}. The $EM$ calculation assumes a constant density in each of the clouds.}
\label{table:1}      
\centering                          
\begin{tabular}{c c c c c}        
\hline\hline                 
Tracer & $T_\mathrm{e}$ & $n_\mathrm{e}$  & Size  & EM  \\   
 & (K) &  (cm$^{-3}$) &  (pc) &  (pc cm$^{-6}$)  \\  
\hline                        
   CRRL & 85 & 0.04 & 35.3, 18.9 & 0.086 \\      
\hline                                   
\end{tabular}
\end{table}

Finally, including the effect of the interstellar medium, we have
\begin{equation}
S_{\nu, \mathrm{measured}} = S_0  \Omega \frac{J(\nu, B)}{4 \pi \chi_\nu (B)} \big( 1 - e^{- \chi_\nu(B) l} \big) \big(f+ (1-f)e^{-\tau_{\mathrm{int}}}\big)e^{-\tau_{\nu, \mathrm{ISM}}}.
\label{full_timeindep}
\end{equation}
The effect of each term on the unabsorbed synchrotron spectrum
is shown in Fig.~\ref{absn_effect}.

\subsection{Secular decline model}

The flux density of Cas A has a time- and frequency-varying secular decline that has been abundantly remarked upon, but remains puzzling. Several studies have attempted to model the time behaviour of Cas A, which is of importance to radio astronomy in general given the long-standing status of Cas A as a calibrator. Recent publications are \cite{helmboldt09, vinyaikin14, trotter17}. The typical procedure is to fit polynomials in log frequency to account for the observed fluctuations in the flux density
of Cas A as a function of frequency and time. The flux density is modelled broadly as $S(t) = S_0 (1 - s)^{t - t_0}$, although with additional terms that try to encompass the frequency dependence of the secular decline. 
The decline rate $s$ has been measured to be 0.9\% yr$^{-1}$ for 1965 \citep{baars77} at 1 GHz, whereas more recently, the decline rate
between 1960 and 2010 has been measured to be an average of 0.6\% yr$^{-1}$ \citep{trotter17} with fluctuations on timescales of years.
At lower frequencies (38 -- 80 MHz), \cite{helmboldt09} find that the secular decrease is stable over five decades with a rate of 0.7\%$-0.8$\% yr$^{-1}$ 
(significantly lower than the value expected from the \cite{baars77} fit at these frequencies, 1.3\% yr$^{-1}$  at 74 MHz). 
All of these papers point out that the secular decline rate varies both over time and with frequency. An important point is that the secular decline rate of Cas A is slightly higher at lower frequencies (i.e. the spectrum of Cas A is flattening). Different frequencies having different values of $s$ implies a change in spectral index over time. 

The models mentioned above provide good fits for the time baseline of around 60 years for which there are radio flux density measurements of Cas A, but are not physically motivated. \cite{shklovsky60}  \citep[see also][]{dubner15} did provide a physical explanation for the secular decline, proposing that it is due to the expansion of the SNR, which causes
the magnetic field to decline as $B \propto R^{-2}$ (from magnetic flux conservation) and the relativistic electrons to adiabatically cool as $E \propto V^{-4/3} \propto R^{-4}$. We do know that Cas A is still actively accelerating electrons \citep[e.g.][]{vink03a,patnaude11}.
For this reason, Shklovskii's model, which only accounts for adiabatic cooling and magnetic flux conservation,  cannot be complete.

A plausible model for the flux decline can be parameterised as  $S(t) = S_0 t\, ^{-\beta}$, which is also
used to model the flux density of radio supernovae \citep[e.g.][]{weiler10}.
The adiabatic expansion of the remnant is not expected
to affect the shape of the electron distribution and therefore 
should have no bearing on the radio spectral index. Hence, adiabatic expansion cannot explain the fact that the decline rate appears to be frequency dependent.

The parameter $\beta$ also connects the well-known (although controversial) $\Sigma-D$ relation with the Sedov evolutionary model. 

According to the  $\Sigma-D$ relation, the diameter $D$ of an SNR goes as a power of its surface brightness $\Sigma$,
\begin{equation}
\begin{split}
D & \propto \Sigma^{-\beta^\prime} \\
& \propto \left( \frac{F_\nu}{D^2} \right)^{\beta^\prime} \\
\end{split}
\label{eqt_sd}
,\end{equation}
that is, $D\,^{2 \beta^\prime + 1} \propto F_\nu \, ^{\beta^\prime}$. Sedov expansion implies that the diameter goes as some power $m$ of time, where $m$ is known as the expansion parameter: $D \propto t^m$. Combining both relations, we arrive at
\begin{equation}
F_\nu \propto t \,^{\frac{m}{\beta^\prime}(2 \beta^\prime +1)}
.\end{equation}
This implies that 
\begin{equation}
\frac{1}{F} \frac{dF}{dt} = \frac{m (2 \beta^\prime +1)}{\beta^\prime} \frac{1}{t}.
\label{eqt_time}
\end{equation}

\subsection{Fitting}

\begin{figure}
   \centering
   \includegraphics[width=9cm]{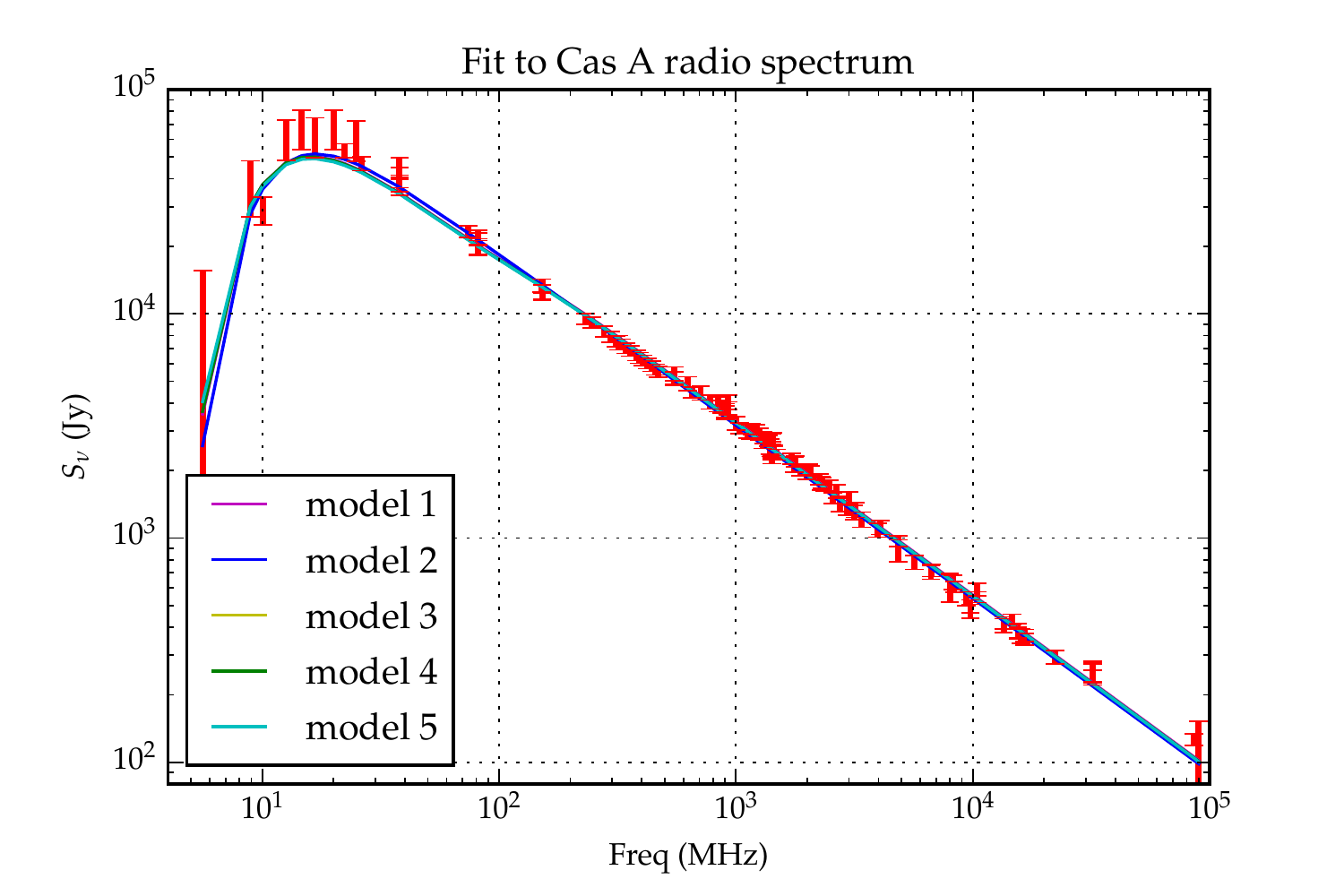}
      \caption{Best-fit models to the Cas A radio spectrum. The data points have been brought to a common epoch using $\beta = 1.80$. The parameters for each model are shown in  Table \ref{table:2}. All models have trouble reproducing the flux density at low frequencies.
              }
         \label{full_spectrum_fit}
   \end{figure}

We compiled a series of radio flux densities from 1960 until 2017\footnote{These are all published points, except for measurements taken between 2015 and 2017 with the Effelsberg single dish, which will be published in Kraus et al., in prep.}, and fitted for the following equation:
\begin{equation}
S_{\nu, t} = S_0  \Omega \frac{J_\nu}{4 \pi \chi_\nu} \big( 1 - e^{- \chi_\nu l} \big) \big(f+ (1-f)e^{-\tau_{\mathrm{int}}}\big)e^{-\tau_{\nu, \mathrm{ISM}}} \left( \frac{t - t_{\mathrm{exp}}}{t_0 - t_{\mathrm{exp}}} \right)^{- \beta}.
\label{full_time_dep}
\end{equation}
We take $t_{\mathrm{exp}} = 1672$ as the time of the explosion  and $t_0 = 1965$ as a reference year. In order to perform the fit, we followed these steps:
\begin{itemize}
\item We fixed $f$ to be the average value inside the reverse shock multiplied by the ratio of the number of unmasked pixels internal to the reverse shock to the total. This gives $f = 0.88$.
\item We either fit for $B$ or fixed it to 0.78 mG (its minimum energy value\footnote{The common reference for the minimum energy value of Cas A is taken from \cite{longair11}, where it is calculated from an outdated distance estimate and for a spherical emitting volume. A change in the distance affects both the luminosity and the size of the emitting volume. With $d=3.4$ kpc and in the case of an emitting shell with $R_{\mathrm{outer}} = 2.5 \arcmin$ and $R_{\mathrm{inner}} = 1.5 \arcmin$, the minimum energy magnetic field is 0.78 mG.}).
\item We fixed the internal emission measure to be our $EM$ average of 37.4 pc cm$^{-6}$.
\item When solving for the ISM component, we set $Z = 1 $. Given the steep dependence of low-frequency absorption with temperature, we assume that the cold phase of the ISM is dominant ($T \sim 20$~K).
\item The terms we fit for are the normalisation constant $S_0$, $\tau_{ISM}$ as parameterised by $EM$ (assuming an ISM temperature of 20 K and $Z = 1$), $\beta$ (which is actually $\beta =  \frac{m (2 \beta^\prime +1)}{\beta^\prime} $ , as shown in Eq. \ref{eqt_time}), and the electron spectral index $p$ (which is related to the radio spectral index $\alpha$ by $p = 2 \alpha +1$).
\item We also made $B$ and $f$ variable terms to fit for.
\item Finally, we also fit using the 0.12 mG lower limit to the magnetic field strength as inferred from gamma rays \citep{abdo10}, since there is no physical
reason to assume that the minimum energy condition holds in Cas A. 
\end{itemize}
The results of these fits are tabulated in Table \ref{table:2} and plotted in Fig. \ref{full_spectrum_fit}. 

\begin{table*}
\caption{Best-fit parameters to Eq. \ref{full_time_dep}. $EM_{\mathrm{int}}$ = 37.4 pc cm$^{-6}$ and $S_{\mathrm{1GHz}}$ corresponds to epoch 1965. The fixed values of $B$ and $f$ were 0.78 mG and 0.88, respectively.}             
\label{table:2}      
\centering          
\begin{tabular}{c c c c c c c c }     
\hline\hline       
Model & $S_{1\mathrm{GHz}}$ (Jy) & $EM_{\mathrm{ISM}}$ (pc cm$^{-6}$) & $p$ & $\beta$ & B (mG) & f & red. $\chi^2$ \\    
\hline                    
1 & 2812 $\pm$ 27 & 0.126 $\pm$ 0.011 & 2.5448 $\pm$ 0.0002 & 1.78 $\pm$ 0.06 & fixed & fixed & 1.49 \\
2 & 2693 $\pm$ 22 &  0.146 $\pm$ 0.013 & 2.5458 $\pm$ 0.0003 & 1.70 $\pm$ 0.07 & fixed & 0.97 $\pm$ 0.02 & 1.34      \\
3 & 2703 $\pm$ 11 & 0.128 $\pm$ 0.012 & 2.5467 $\pm$ 0.0004 & 1.79 $\pm$ 0.08 &  0.737 $\pm$ 0.008 & fixed & 1.49  \\
4 & 2603 $\pm$ 30 & 0.146 $\pm$ 0.013 & 2.5438 $\pm$ 0.0005 & 1.80 $\pm$ 0.08 & 0.691 $\pm$ 0.011 & 0.95 $\pm$ 0.02 & 1.32 \\
5 & 2728 $\pm$ 21 & 0.142 $\pm$ 0.012 & 2.5460 $\pm$ 0.0001 & 1.72 $\pm$ 0.07 & 0.12 (fixed) & fixed & 1.75 \\
\hline                  
\end{tabular}
\end{table*}

\subsection{Comparison with the $\Sigma-D$ relation and SN 1993J}

The $EM$ values due to ISM absorption are consistently around 0.13 pc cm$^{-6}$. These are higher than the $EM$ value due to the Perseus arm components found by \cite{oonk17}. Part of the
reason for this is that they did not model the CCRL component in the Orion arm, and an additional component with a higher $n_\mathrm{e}$ might be present, as is suggested by hydrogen recombination lines in the same work. In fact, our measurement of $EM$ provides an upper limit on the average electron density along the line of
sight. Around two-thirds of the free electrons along the line of sight are associated with the cold neutral medium clouds that CRRLs trace. This suggests that at low frequencies we do not have a smooth, absorbing electron medium, but rather a clumpy one. 

It is difficult in general, and not possible from our measurements, to distinguish the effects of synchrotron self-absorption and free-free absorption from the ISM. Separating them depends on the lowest frequency data points. These are the most unreliable in our data set, which is due both to the ionospheric cutoff of the radio window that occurs at around 10 MHz, and also to the fact that the secular decline at the lowest frequencies is poorly understood. We explore the latter point  in the next section.

Our values of $\beta$ vary between 1.70 and 1.80. These imply a power-law index for the $\Sigma - D$ relation (see Eq. \ref{eqt_sd}) of $\beta^\prime = 1.74$ and $\beta^\prime = 1.37,$ respectively, for an expansion parameter $m = 0.66$ \cite[]{patnaude09}. This is very different from the power-law index recently found for the Large Magellanic Cloud of $\beta \sim 3.8$ \cite[]{bozzetto17}, which is comparable with other nearby galaxies (these are more reliable than the values in our Galaxy, where the distances to SNRs are poorly determined, and hence so are their diameters). However, the $\Sigma - D$ relation has a notoriously large scatter, and particularities in the environment of Cas A such as the fact that it is evolving within the cavity of a stellar wind can account for this discrepancy.

A caveat of this temporal model is that it cannot have held for the entire lifetime of Cas A, or it would have been too bright at the time of explosion. 
Taking $\beta=1.8$ and extrapolating the current radio luminosity  of Cas A back to $\sim 1673$, one year after the explosion,
gives a radio luminosity of  $L_{1 \rm{GHz}} = 1.4 \times 10^{30}$ erg cm$^{-2}$s$^{-1}$ Hz$^{-1}$. 
For comparison, SN 1993J (the prototype for a Type IIb SN) had  $L_{1.5 \rm{GHz}} = 1.6 \times 10^{27}$ erg cm$^{-2}$s$^{-1}$ Hz$^{-1}$\cite[]{weiler10}.
The Cas A and SN 1993J supernovae were, at least in the optical, very similar \cite[]{krause08}, and it is unlikely that the radio luminosity of Cas A was three orders of magnitude higher than that of 1993J. 
Models where the flux density varies as an exponential with time do not have this problem. The secular decline reported in the classical work of \cite{baars77} would imply that their 2723 Jy flux at 1 GHz in 1965 was actually 37\,500 Jy after the explosion. Even for their very high value of the secular decline rate, the luminosity corresponds to only $L_{1 \rm{GHz}} = 5.2 \times 10^{26}$ erg cm$^{-2} $s$^{-1}$ Hz$^{-1}$.

For SN 1993J, the flux decline was characterised by a decline parameter $\beta\approx 0.7$, which is much smaller than Cas A's. 
If the radio luminosity of Cas A in its initial years was similar to SN1993J with $\beta\approx 0.7$, Cas A must have rebrightened, and
 have had a faster decline with $\beta \approx 1.8$ since. 
There are several documented examples of supernovae that have rebrightened in the radio, such as \cite{corsi14, salas13, soderberg06}.
We speculate that in the case of Cas A, such a rebrightening may have been caused by the blast wave hitting a density enhancement
in the wind of the progenitor, or it may have been related to an increased radio luminosity from the formation of the bright ring, which probably
corresponds to shocked ejecta \citep{helder08}. The formation of the ring may in fact have been caused by a sudden deceleration of
the forward shock, which increases the velocity with which the ejecta are being shocked.

\subsection{Frequency dependence of the secular decline}

It is evident from Fig. \ref{full_spectrum_fit} that all models have trouble reproducing the flux density at low frequencies, and that there is some component that remains to be modelled. We propose two effects that can be responsible for this discrepancy: the existence of two electron populations with a different spectral index and a different secular decline rate, or having an amount of unshocked ejecta that varies with time.

\vspace{0.5cm}
\noindent
\textbf{Two electron populations} 

Radio supernovae have steep spectral indices of $\alpha \sim 1$, whereas older SNRs tend to have flatter values, $\sim$ 0.5--0.55. If supernova remnants have multiple electron populations, one or the other could dominate the radio emission at earlier versus later times. Cas A is young and has
an intermediately steep radio spectrum ($\alpha=0.77$). This
means that it might be in a transitional phase where both electron populations contribute significantly to the radio spectrum, but they decline differently, resulting in a change of spectrum with time. 

Adiabatic cooling and weakening of the magnetic field due to expansion are the most important processes that change the emissivity
of old electron populations, but they do not alter the spectral index.
The spectral index can change if newly accelerated electrons have an inherently different power law distribution, and/or if there are two different electron
populations with different spectral indices $\alpha$ and different secular decline parameters $\beta$.

We can model this situation assuming 
\begin{equation}
\begin{split}
S_{\nu, t} &= \left( A_1\left( \frac{\nu}{\nu_0} \right)^{-\alpha_1} \left( \frac{t - t_{\mathrm{exp}}}{t_0 - t_{\mathrm{exp}}} \right)^{- \beta^\prime_1} + A_2 \left( \frac{\nu}{\nu_0} \right)^{-\alpha_2} \left( \frac{t - t_{\mathrm{exp}}}{t_0 - t_\mathrm{{exp}}} \right)^{- \beta^\prime_2} \right) \\
& \times \big(f+ (1-f)e^{-\tau_{\mathrm{int}}}\big)e^{-\tau_{\nu, \mathrm{ISM}}}.
\end{split}
\end{equation}
We fitted for this equation and found values of 
$\alpha_1=0.7546 \pm 0.0009$,  $\alpha_2=0.7821 \pm 0.0006$, $\beta_1 = 0.6 \pm 5.1,$ and $\beta_2 = 2.2 \pm 1.6$, with a flux density ratio of the two components of $A_2/A_1=1.04$, and a reduced $\chi^2 = 1.5$. The $EM_{\mathrm{ISM}}$ value is $0.137 \pm 0.012$, similar to the best fit four our other models. 

The improvement to the fit is marginal $\Delta \chi^2 = 6$ and the decline rate parameters $\beta_{1,2}$ are ill-constrained. Nevertheless, this may be a
possible solution to the frequency dependence of the flux density decline. It could be verified by spatially identifying regions of different spectral index and
measuring their flux density decline. In practice, this requires decades of mapping the source with high flux density accuracies.

\vspace{0.2cm}
\noindent
\textbf{Time-varying internal absorption from the unshocked ejecta} 

Another source of time-varying effects in Eq. \ref{full_time_dep} could come from the term of internal absorption $f - (1-f) e^{-\tau_\nu}$. 
If $\tau_{\nu, \mathrm{int}}$ varies with time, absorption can have a different impact on the measured flux density at different times, and this can appear as a variation with frequency of the secular decline rate instead of as a variation with time of the amount of absorption present at any given frequency (for a synchrotron flux decaying at the same rate throughout the frequency range). 

We know that the cold ejecta are continuously encountering the reverse shock and heating up. 
This implies that, at earlier times, there was more cold mass of higher densities that could absorb at low frequencies. 
In general, this would have the effect of steepening the Cas A flux density (it would look like the flux is decaying more slowly at lower frequencies), which is the opposite of what is observed. 

Here we discuss what time dependency the decreasing amount of cold mass can have on the internal absorption coefficient $\tau_{\nu, \mathrm{int}}$. 
Assume that the free-free optical depth in fact goes as
\begin{equation}
\tau_{\mathrm{int}} (\nu, t) \propto \left(\frac{\nu}{\nu_0}\right)^{-2} \, Z\, T^{-3/2} \, EM_0 \, t^{-\xi}.
\end{equation}
For free expansion, the density of the gas scales with time as $\rho \propto t^{-3}$ \cite[]{chevalier82}.
If the absorption is due to diffuse unshocked gas, the emission measure would scale as $EM \propto l t^{-6}$,
with $l$ the absorption length scale.
If the reverse-shock radius does not change, this implies $\xi=6$.
However, if the absorption is due to discrete dense clumps, it is the density of clumps that changes as $t^{-3}$, which implies,
for a fixed $l$, $\xi=3$.

In hydrodynamical models of Cas A \citep[e.g.][]{orlando16}, the reverse shock is still moving outward, which implies a lower value of $\xi$.
However, optical\footnote{As presented by R. Fesen at the CSI workshop at Princeton in 2017 \url{http://www.kaltura.com/index.php/extwidget/preview/partner_id/1449362/uiconf_id/25928631/embed/auto?&flashvars[streamerType]=auto&flashvars[playlistAPI.kpl0Id]=1_qps3id8h}.} 
and X-ray measurements in the west \citep[see the discussion in][]{helder08} show that at least in some regions, the reverse shock is close to
a stand-still, and $\xi=3$ or $\xi=6$ for a model with clumping or diffuse ejecta, respectively.

\begin{figure}
   \centering
   \includegraphics[width=9cm]{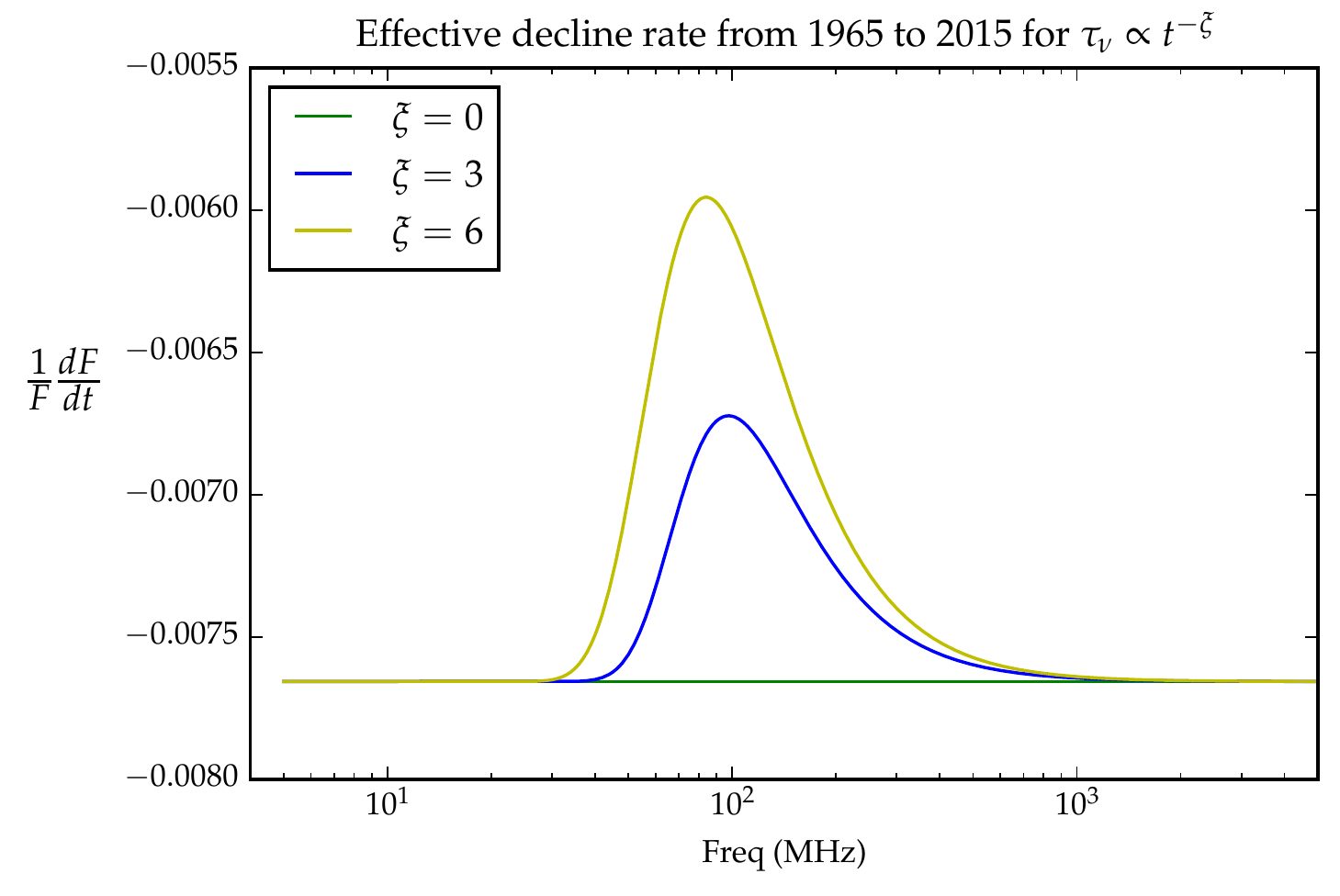}
      \caption{Effective decline rate for a 50-year time difference in units of percentage points per year for a source whose synchrotron flux varies uniformly across the frequency range as $t^{-2}$ but that has less low-frequency absorption at later times. The \lq effective' decline rate is lower at a range of low frequencies around 10 MHz.
              }
         \label{eff_decline}
   \end{figure}

Figure~\ref{eff_decline} shows the \lq effective decline rate' $\frac{1}{F}\frac{dF}{dt}$ for a time-varying $\tau_{\mathrm{int}}$ in Eq. \ref{eff_decline}. The \lq effective' decline rate is lowest at a range of low frequencies around 100 MHz. The position left-to-right of the peak in the graph is determined by the values of $EM_{\mathrm{int}}$, and its vertical scale is determined by $\xi$. 

In this plot the flux density is taken to result from synchrotron emission that varies uniformly across the frequency range as $t^{-2}$. The effect of a smaller amount of low-frequency absorption at later times appears as an \lq effective' frequency-dependence in the secular decline rate. In this model we did not include the effect of change in the absorbed fraction $f$, which can be of a similar order as that of the changing $EM$, but depends on the particulars of the geometry of the unshocked ejecta, for which we cannot account. 

The effect on the decline rate may be quite considerable, 10\%, and it has a distinct frequency dependence that could be measurable.
The fact that the decline rate seems to increase, rather than decrease at lower frequency, means that changes due to internal absorption
are relative small, and more consistent with a low value of $\xi$. This further strengthens the suggestion that the unshocked ejecta might be in the form of clumps. 
The  effect of changes in the internal absorption over decades of observations could be important and can take place alongside the effect described above of multiple electron populations with different power-law indices and decay rates. If these effects occur in concert, they can give rise to a complex flux density evolution with time at lower frequencies.


\section{Conclusions}

We have imaged Cas A with the \lofar\ LBA, creating a range of maps from 30 to 77 MHz.  We have used these, along with VLA images at 330 MHz, 1.4 GHz and 5 GHz, to fit for
free-free absorption from cold gas inside the reverse shock of Cas A. 
In addition, we used published flux density measurements of Cas A with a broad time and frequency baseline to understand the
effect the internal cold gas has on the integrated radio spectrum of Cas A, including its time evolution.
We summarise our results as follows:

   \begin{enumerate}
      \item At low frequencies, the area internal to the reverse shock shows clear absorption features, with regions of high departure from power-law behaviour. 
      On average, we measure that 78\% of the synchrotron emission from the projected area of the reverse shock comes from the front side of the shell. For a temperature of $T=100$ K and an average ionisation state of $Z=3,$
      we measure an average emission measure of $EM = 37.4 \, \rm{pc} \, \rm{cm}^{-6}$.
      \item For these same gas conditions and a geometry where the mass is in a sheet 0.16 pc thick, our mass estimate in the unshocked ejecta is $M = 2.95 \pm {0.48}$ \msun, which is quite high given our knowledge about the supernova progenitor and the current dynamical state of Cas A.
      \item If the unshocked ejecta are clumped, the mass can be reduced significantly and still be responsible for the low-frequency absorption.
      \item If the gas temperature is lower than 100 K, the mass estimate can also be reduced. We find it likely that the unshocked ejecta are colder than 100 K,
      but detailed non-equilibrium modelling of the infrared flux is necessary to verify this. 
      \item We measure the reverse shock to have a radius of $114$\arcsec $\pm $6\arcsec\ and be centred at 23:23:26, +58:48:54 (J2000). We find the radio reverse shock is at a larger radius and more centrally located than the reverse shock as probed by non-thermal X-ray filaments. 
      \item We measure the ISM absorption along the line of sight to Cas A to be $EM = 0.13 \, \rm{pc} \, \rm{cm}^{-6}$ for a 20 K ISM. 
      \item We explore the effects that Cas A having two electron populations and having a time-varying mass in the unshocked ejecta could have on the secular decline.
      These effect are competing, and their combination could be responsible for the frequency dependency of the secular decline, which is not explained if adiabatic expansion is responsible for 
       declining flux density of Cas A, as well as for its counterintuitive, varying temporal behaviour.
   \end{enumerate}

\begin{acknowledgements}

We thank T. Delaney for her VLA images of Cas A, and R. Perley and A. Kraus for making their recent flux density measurements with the VLA and Effelsberg available to us. We also thank the internal referee from the \lofar\ builder' list and the anonymous referee from A\&A. Their helpful comments and suggestions improved the quality of this paper.

The work of MA and JV is supported  by a grant from the Netherlands Research School for Astronomy (NOVA). PS and JBRO acknowledge financial support from the Dutch Science Organisation (NWO) through TOP grant 614.001.351. RJvW acknowledges support from the ERC Advanced Investigator programme NewClusters 321271. 

This paper is based (in part) on data obtained with the International \lofar \, Telescope (ILT). \lofar \, \cite[]{vanhaarlem13} is the Low Frequency Array designed and constructed by ASTRON. It has facilities in several countries, that are owned by various parties (each with their own funding sources), and that are collectively operated by the ILT foundation under a joint scientific policy. \lofar\ data reduction used the DRAGNET GPU cluster (at the CIT in Groningen), which was funded by the European Research Council under the European Union's Seventh Framework Programme (FP7/2007-2013) / ERC grant agreement nr. 337062 (PI: Hessels). The LOFAR software and dedicated reduction packages on \url{https://github.com/apmechev/GRID_LRT} were deployed on the e-infrastructure by the LOFAR e-infragroup, consisting of R. Oonk (ASTRON \& Leiden Observatory), A. P. Mechev (Leiden Observatory) and T. Shimwell (Leiden Observatory) with support from N. Danezi (SURFsara) and C. Schrijvers (SURFsara). This work has made use of the Dutch national e-infrastructure with the support of SURF Cooperative through grant e-infra160022  \& 160152. The National Radio Astronomy Observatory is a facility of the National Science Foundation operated under cooperative agreement by Associated Universities, Inc. 
\end{acknowledgements}

\begin{appendix}\section{\lofar\ LBA images}

We present as an appendix the gallery of \lofar\ images that are the basis for this work. 
\begin{figure*}
\centering
\includegraphics[width=0.84\paperwidth]{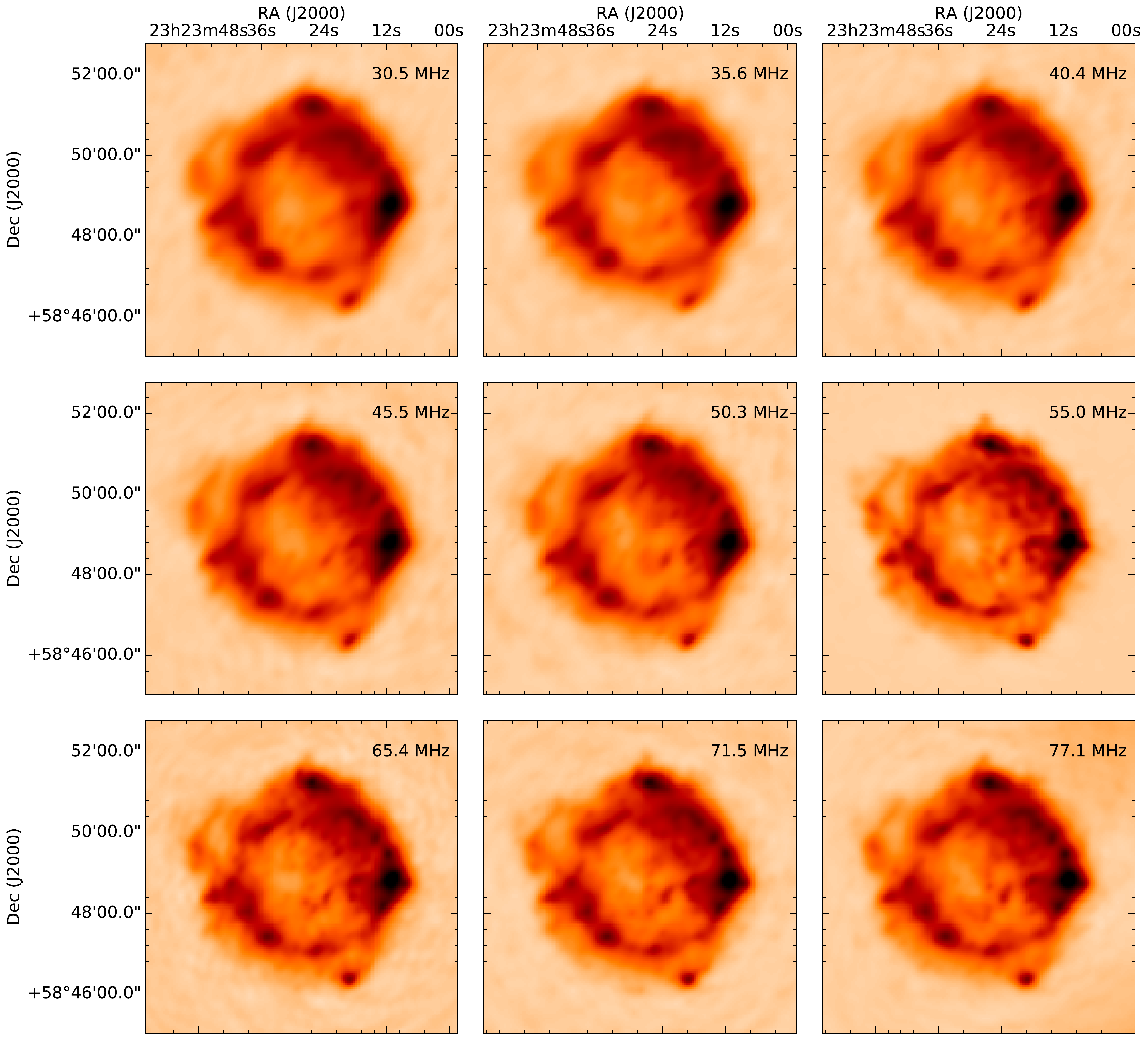}
\caption{Gallery of narrow band \lofar\ LBA images. For each map, the bandwidth is 1 MHz with central frequencies as indicated by the wedges. All have a common $u-v$ range of 500 --12\,000 $\lambda$, which corresponds to synthesised beam sizes of  17\arcsec to 7\arcmin.}
\label{appfig}
\end{figure*}

\end{appendix}

\end{document}